# A disease-contextualized atlas of proteome-scale molecular influences of natural products

Boyang Wang[1,#], Jun Kang[2], Bo Zhang[1,#]

[1] Sino-Glow Intelligent Link (Tianjin) Life Science and Technology Co., Ltd., Nankai District, Tianjin 300192, China.
[2] School of Life Sciences, Tianjin University, 92 Weijin Road, Nankai District, Tianjin 300072, P.R. China.

[#] Correspondence to: boyang.hyzt@gmail.com (Boyang Wang) and zhangbo.2007@tsinghua.org.cn (Bo Zhang)

**Abstract**
Understanding the system-wide molecular effects of natural products remains challenging because of fragmented chemical knowledge, incomplete proteomic coverage and limited experimental perturbation data, particularly for multicomponent medicinal herbs. Here, we present an integrative computational framework for modeling the disease-contextualized molecular influences of natural compounds and medicinal herbs. Through structure-based harmonization of three major herbal medicine databases, we established a standardized atlas comprising 46,422 chemical entities, 12,459 medicinal herbs and 156,588 herb–compound associations. We further developed a protein-centered strategy integrating high-confidence protein–protein interactions with proteomic and transcriptomic data to reconstruct disease-specific molecular networks across nine cancer types. By leveraging transcriptomic associations to complement incomplete proteomic coverage, these networks captured disease-associated molecular rewiring beyond conventional differential expression. Coupling artificial intelligence inferred compound–target interactions with signed random walks with restart enabled the modeling of potential molecular influence propagation, extending sparse direct-target associations into directional, genome-scale molecular influence profiles. Using gastric cancer as an in-depth case study, we generated network-wide influence profiles for 45,418 compounds across 16,899 molecular nodes and extended the approach to multicomponent herbal interventions. We further applied the framework to eight additional cancer types, demonstrating its applicability to diverse disease-specific molecular contexts. Retrospective proteomic perturbation analyses provided preliminary support for the biological relevance of these inferred influences. Together, our study establishes a scalable framework connecting natural-product chemical knowledge with disease-specific molecular organization, providing a foundation for systems-level modeling of natural-product interventions and mechanism-guided therapeutic discovery.

## Introduction

Artificial intelligence (AI) has emerged as a transformative force in biomedical research, enabling a shift from descriptive analyses of biological data toward predictive and generative modeling of molecular functions, cellular processes, and therapeutic responses[1]. At the molecular level, advances in deep learning have substantially improved the prediction of protein structures and biomolecular interactions, as exemplified by AlphaFold, ESMFold, and AlphaFold 3[2-4]. Generative models have further extended these capabilities to the de novo design of proteins with desired structural and functional properties[5]. At the genomic and cellular levels, AI approaches have enabled increasingly sophisticated analyses of gene regulation, cellular states, and context-dependent molecular networks[6-8]. These advances have also accelerated therapeutic discovery, facilitating the identification of novel bioactive compounds and the development of AI-designed drug candidates, some of which have progressed into clinical evaluation[2,5,9]. Collectively, these developments highlight the growing potential of AI to bridge molecular-level predictions and disease-specific biological contexts, laying the foundation for understanding therapeutic actions within interconnected molecular systems.

Natural products, encompassing chemically diverse bioactive molecules derived from plants, microorganisms, and other organisms, have long served as indispensable sources of therapeutic agents, particularly for cancer and infectious diseases[2,5,9]. Beyond individual bioactive compounds, medicinal herbs represent complex multicomponent systems containing diverse chemical constituents that may interact with multiple molecular targets and biological pathways, offering opportunities to investigate therapeutic effects beyond single-target pharmacology[10,11]. Recent advances in artificial intelligence and multi-omics technologies have further transformed natural product research, enabling large-scale chemical characterization, bioactivity screening, target identification, and mechanism-of-action investigation[12,13]. At the molecular level, foundation models tailored to natural product chemistry have improved molecular representation learning and facilitated the identification of bioactive candidates[14], while generative approaches have enabled the design of novel natural-product-inspired molecules with experimentally validated biological activities[15]. Beyond molecular activity prediction, AI-driven perturbation modeling has extended these capabilities toward genome-wide characterization of therapeutic responses. For example, PRnet employs deep generative learning to predict transcriptional responses to previously unseen compounds and supports the identification of anticancer natural products[16]. Similarly, SETComp integrates transfer learning and permutation-invariant representations to predict genome-wide, cell-specific, and directional transcriptional responses to multicomponent interventions[17]. Together, these developments demonstrate the growing potential of AI to connect the chemical diversity of natural products and the compositional complexity of medicinal herbs with their molecular interactions and cellular responses, providing new opportunities for investigating their therapeutic effects at a systems level.

Despite these advances, several fundamental challenges continue to limit the application of AI to natural product research. First, experimentally validated natural product–target interactions and perturbation profiles remain scarce and fragmented, necessitating substantial reliance on knowledge transfer from datasets dominated by conventional small-molecule drugs. Differences in chemical space, together with biased interaction annotations, can compromise the generalizability of existing models to structurally diverse natural products[14,18]. Second, although proteins constitute the principal molecular targets of most therapeutic drugs, existing approaches often focus either on direct drug–target interactions with incomplete or unevenly validated proteome-wide coverage, or on predicting transcriptional responses to chemical perturbations[16,19]. Importantly, transcriptomic alterations do not necessarily reflect corresponding changes in protein abundance or function, highlighting the need to incorporate proteomic information when characterizing disease-associated molecular states[20,21]. Third, most AI models are developed for individual compounds, whereas comparatively few approaches explicitly address the compositional complexity of medicinal herbs. Although recent efforts have extended intervention prediction to multicomponent systems, herb-level modeling remains constrained by limited perturbation data and heterogeneous compound annotations across databases[17,22]. Finally, molecular binding predictions and transcriptional perturbation signatures are often disconnected from disease-specific molecular interaction networks, limiting the ability to characterize how direct target engagement may influence broader biological systems through interconnected protein interactions[23]. Together, these limitations underscore the need for an integrative framework that connects standardized natural product resources, proteome-wide target inference, and multi-omics-informed disease-specific networks to systematically characterize the potential molecular effects of individual compounds and complex herbal systems.

Network-based approaches offer a promising strategy for addressing data scarcity in therapeutic discovery by leveraging prior knowledge of molecular interactions and biological organization. By integrating heterogeneous biological relationships, graph-based models can facilitate knowledge transfer across related molecular entities and disease contexts, supporting therapeutic predictions under few-shot and, in appropriately designed models, zero-shot settings[24,25]. Moreover, the explicit representation of molecular interactions and network topology provides a biologically interpretable framework for tracing potential relationships between drug targets, disease-associated proteins, and functional modules[26,27]. Together, these advances suggest that integrating compound–protein interactions with disease-specific molecular networks could provide a scalable and interpretable approach to modeling the potential proteome-wide effects of natural products, including complex herbal systems.

Here, we developed an integrative computational framework that connects natural product chemical resources, high-confidence protein interaction networks, and multi-omics data to characterize disease-specific molecular states and predict network-wide therapeutic effects. We first systematically harmonized multiple herbal medicine databases through standardized chemical annotation and PubChem-based compound resolution, establishing a unified resource of medicinal herbs, natural compounds, and herb–compound associations. In parallel, we integrated multiple protein–protein interaction databases with transcriptomic and proteomic profiles to reconstruct disease-specific molecular networks across nine cancer types, enabling systematic characterization of cancer-associated network remodeling and molecular interaction rewiring. Building on these networks, we combined AI-based drug–target interaction prediction with signed random walks with restart (SRWR) to propagate compound-associated molecular signals from predicted protein targets across disease-specific interaction networks, generating directional, proteome-wide network influence profiles. We further extended this strategy from individual natural compounds to medicinal herbs by integrating the network-level influence profiles of their constituent compounds, and evaluated its biological relevance using independent molecular perturbation evidence and disease-relevant case studies. Collectively, our study provides standardized herb–compound and protein interaction resources alongside a generalizable network-based framework for investigating the potential system-level effects of natural products in defined disease contexts, bridging molecular target prediction and the systems-level characterization of complex therapeutic interventions.

## Results

### Construction of a unified natural-product knowledge atlas and an integrative network-based modeling framework

To enable systematic characterization of natural-product effects within disease-specific molecular contexts, we established an integrative computational framework connecting standardized herbal medicine resources, multi-omics-informed molecular networks, and AI-guided network propagation (Fig. 1). The framework comprises four interconnected components: (i) harmonization of herbal medicines, natural compounds, and their compositional relationships; (ii) reconstruction of cancer-specific molecular interaction networks through the integration of transcriptomic, proteomic, and protein–protein interaction data; (iii) inference of compound–protein interactions followed by SRWR to generate disease-contextualized, proteome-wide influence profiles; and (iv) extension of compound-level predictions to multicomponent herbal systems. By establishing a shared molecular representation across these components, the framework was designed to connect the chemical identities and compositions of natural products with their potential effects on disease-associated molecular networks.

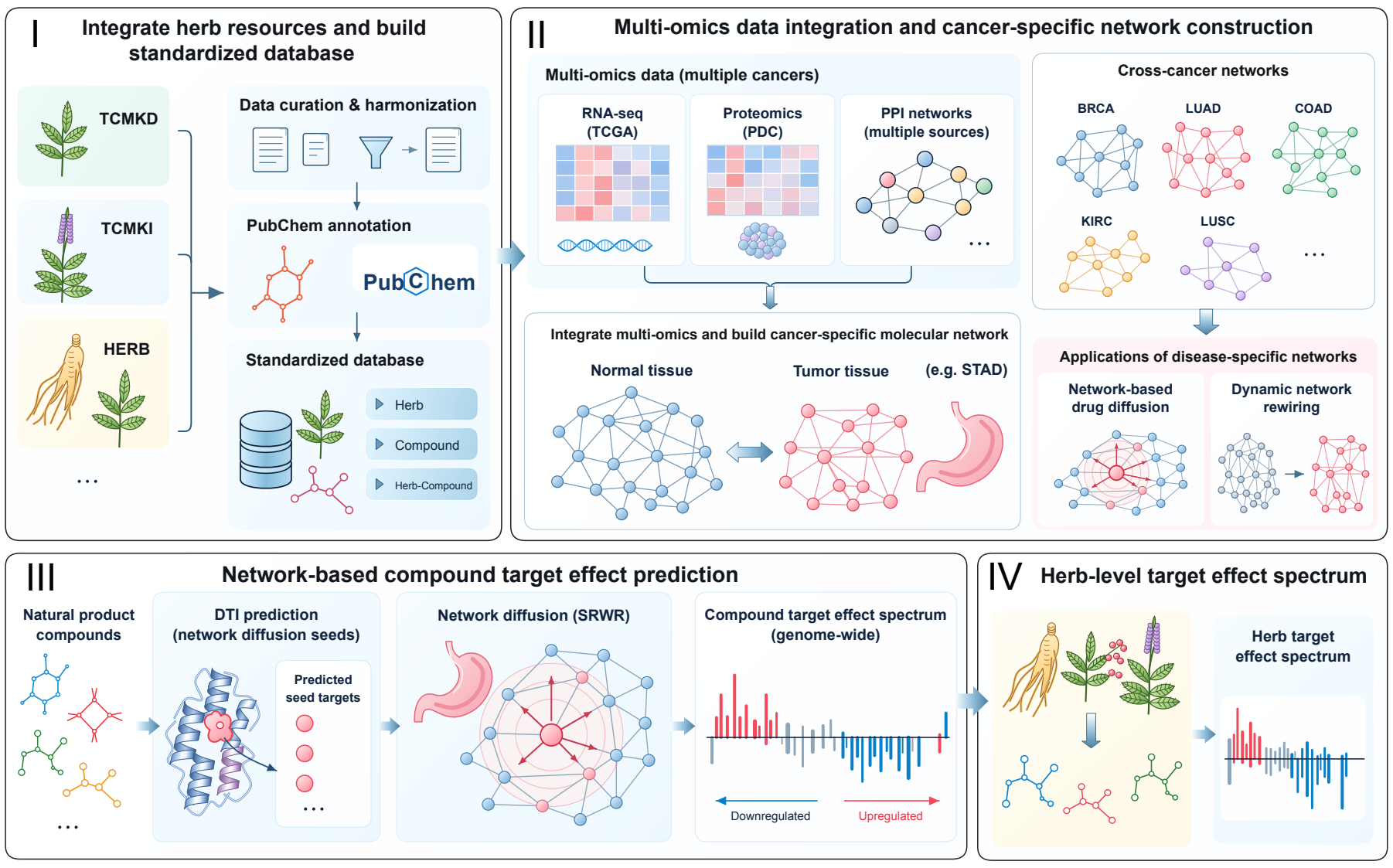

**Fig. 1 Overview of the integrative computational framework for disease-contextualized modeling of natural-product interventions. (I)** Construction of a standardized herb–compound knowledge atlas through systematic integration, chemical harmonization, and PubChem-based annotation of multiple herbal medicine databases, establishing unified herb identities, chemical entities, and herb–compound associations. **(II)** Reconstruction of cancer-specific molecular interaction networks by integrating high-confidence protein–protein interactions with transcriptomic and proteomic data. Protein-centered integration incorporates transcriptomic associations to extend molecular coverage, enabling the characterization of disease-associated network rewiring and providing disease-specific networks for downstream perturbation modeling. **(III)** Modeling of compound-associated molecular influence through AI-based drug–target interaction inference and signed random walks with restart (SRWR). Predicted compound–protein interactions and regulatory-direction information serve as initial propagation signals, generating directional, genome-scale molecular influence profiles within disease-specific networks. **(IV)** Extension of compound-level influence modeling to multicomponent medicinal herbs by integrating the network influence profiles of their constituent compounds, enabling the characterization of potential herb-level molecular intervention spectra.

We first integrated three large-scale herbal medicine databases, HERB 2.0, TCMKI, and TCMKD, which in total comprised 162,008 original ingredient records (Fig. 2A). Because the source databases differed substantially in chemical annotations and identifier completeness, we implemented a hierarchical, structure-first chemical resolution strategy that prioritized molecular identity over compound-name similarity. Existing PubChem compound identifiers (CIDs) were cross-checked against available structural evidence, whereas records lacking reliable identifiers underwent additional structure-based resolution or conservative name- and CAS-supported rescue. This procedure yielded 110,313 PubChem-resolved source records, corresponding to 46,423 unique CIDs. Further examination of 66 groups sharing identical full InChIKeys but containing multiple PubChem CIDs revealed that only one CID pair satisfied our exact stereochemistry-aware molecular graph equivalence criterion. After conservative identity reconciliation, we established a unified registry of 46,422 non-redundant, graph-defined chemical identities, each assigned a stable internal identifier while retaining its original source annotations and PubChem cross-references.

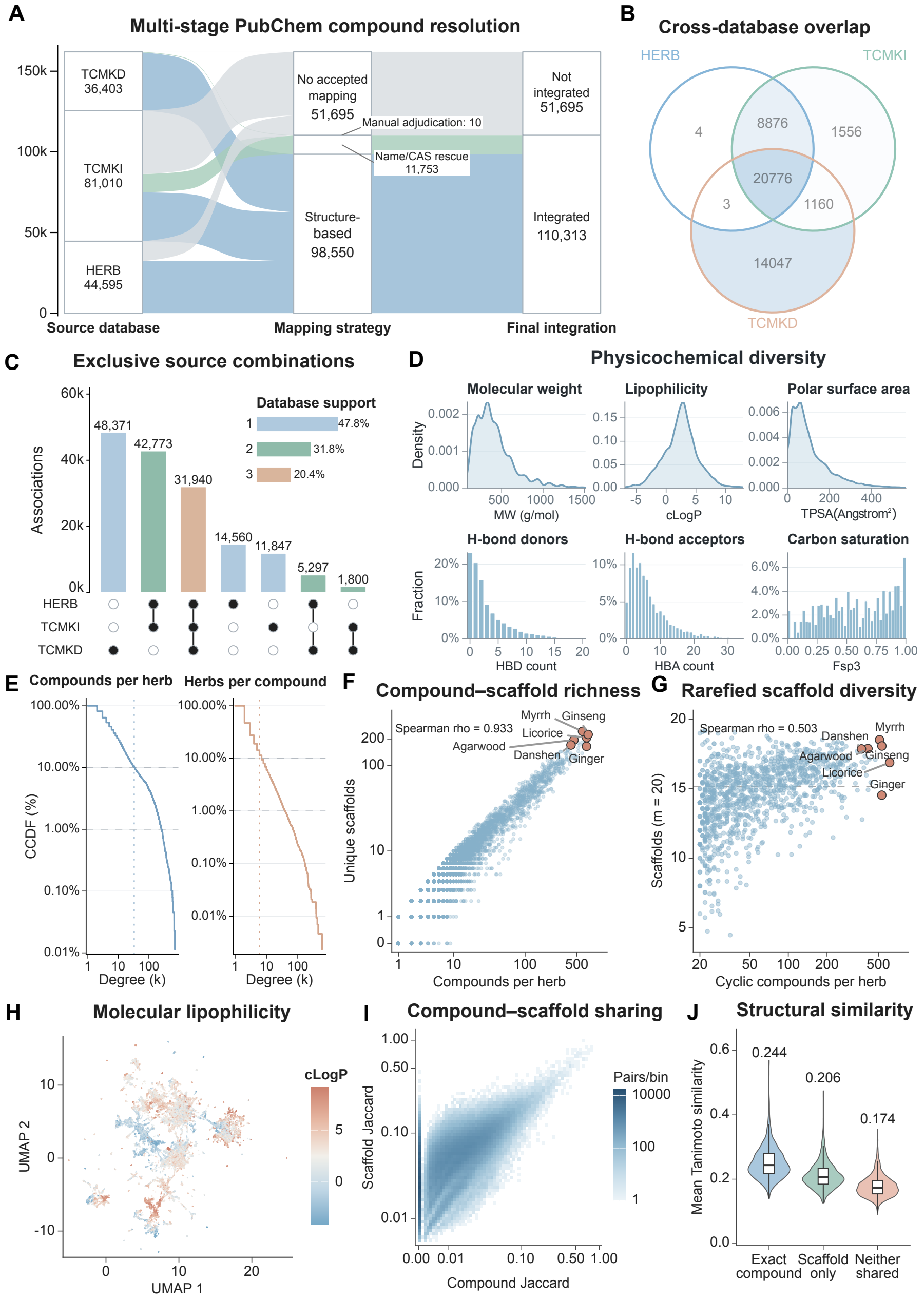

**Fig. 2 Construction and chemical characterization of the standardized herb–compound knowledge atlas. (A)** Multi-stage chemical identity resolution of ingredient records from HERB 2.0, TCMKI and TCMKD. The Sankey

diagram illustrates structure-based PubChem annotation, conservative name- and CAS-assisted resolution, and the retention or exclusion of source records during integration. **(B)** Venn diagram showing the overlap of standardized chemical entities among the three source databases following cross-database identity reconciliation. **(C)** UpSet plot showing the distribution of unique herb–compound associations across exclusive combinations of source databases. Filled circles indicate database membership, and bar colors distinguish associations supported by one, two or all three databases. **(D)** Physicochemical diversity of standardized compounds, including molecular weight (MW), calculated lipophilicity (cLogP), topological polar surface area (TPSA), hydrogen-bond donor and acceptor counts, and the fraction of $sp^3$-hybridized carbon atoms ($Fsp^3$). **(E)** Complementary cumulative distributions of the number of compounds associated with each herb and the number of herbs containing each compound, illustrating the heterogeneity of herb composition and compound sharing. **(F)** Relationship between the number of compounds per herb and molecular scaffold richness, defined as the number of distinct scaffolds represented within each herb (Spearman's $\rho = 0.933$). **(G)** Rarefied scaffold diversity, estimated by standardizing the sampling depth to 20 cyclic compounds per herb, plotted against the total number of cyclic compounds associated with each herb (Spearman's $\rho = 0.503$). **(H)** Uniform manifold approximation and projection (UMAP) of the standardized chemical space, with compounds colored according to cLogP. **(I)** Relationship between compound-level and scaffold-level compositional overlap across herb pairs, quantified using Jaccard similarity. Color intensity represents the density of herb pairs. **(J)** Distributions of molecular fingerprint-based structural similarity between herb pairs sharing at least one identical compound, sharing molecular scaffolds but no identical compounds, or sharing neither. Violin plots illustrate the distributions of mean pairwise Tanimoto similarity, with internal boxplots summarizing their central tendency and dispersion.

Cross-database comparison revealed both substantial overlap and pronounced source-specific contributions to the integrated chemical space (Fig. 2B). Among the 46,422 graph-defined chemical identities, 20,776 were represented in all three databases, whereas 14,047 and 1,556 were exclusive to TCMKD and TCMKI, respectively. HERB and TCMKI exhibited particularly extensive overlap, indicating that cross-database concordance should not necessarily be interpreted as independent chemical evidence. These observations underscore the value of combining complementary resources while preserving source provenance and distinguishing shared molecular identities from redundant annotations.

We next reconstructed and harmonized the herb–compound relationships associated with the standardized chemical identities. Conservative integration of 23,876 source herb-entity and herb-name nodes yielded 12,459 unified herb identities, including 6,862 represented in multiple databases. Mapping the original herb–ingredient relationships to the standardized compound registry yielded 270,647 source-local herb–compound associations, which were consolidated into 156,588 unique herb–compound relationships after herb identity harmonization (Fig. 2C). Of these, 81,810 (52.2%) were annotated in at least two source databases, whereas 31,940 (20.4%) were represented in all three. Importantly, uncertain herb names and potentially distinct medicinal materials were retained rather than forcibly merged, thereby preserving compositional provenance while reducing redundant relationships.

The resulting atlas encompassed broad physicochemical and structural diversity (Fig. 2D–J). Standardized compounds spanned heterogeneous distributions of molecular weight, lipophilicity, polar surface area, hydrogen-bonding capacity, and carbon saturation (Fig. 2D), with marked differences in the number of constituents associated with individual herbs and the number of herbs sharing a given compound (Fig. 2E). Chemical scaffold richness increased strongly with the number of compounds per herb (Spearman's $\rho = 0.933$; Fig. 2F). However, scaffold diversity remained heterogeneous after rarefaction to 20 cyclic compounds per herb, with a more moderate association between rarefied diversity and constituent number (Spearman's $\rho = 0.503$; Fig. 2G), indicating that differences in herbal chemical diversity were not captured by raw constituent counts alone. Low-dimensional visualization further revealed a heterogeneous chemical landscape with substantial variation in molecular lipophilicity (Fig. 2H). Comparisons of herb pairs showed that compositional overlap and scaffold sharing captured related but nonidentical dimensions of chemical similarity (Fig. 2I), while mean structural similarity progressively decreased across pairs sharing exact compounds, sharing only molecular scaffolds, or sharing neither (0.244, 0.206, and 0.174, respectively; Fig. 2J).

Collectively, these analyses establish a chemically resolved, provenance-preserving, and compositionally diverse natural-product knowledge atlas comprising 46,422 standardized compounds, 12,459 medicinal herb identities, and 156,588 unique herb–compound associations. Beyond providing a consolidated chemical resource, the atlas creates a consistent connection between individual compounds and complex herbal systems, establishing the data foundation for subsequent AI-based target inference and disease-specific molecular network modeling.

**Multi-omics integration enables the reconstruction of cancer-specific molecular networks**

To capture the molecular organization of cancer beyond individual changes in gene or protein abundance, we developed a multi-omics network reconstruction strategy integrating transcriptomic and proteomic profiles with experimentally supported protein–protein interaction (PPI) relationships. We assembled cancer-associated molecular datasets from The Cancer Genome Atlas (TCGA), the Genotype-Tissue Expression (GTEx) project, and the Proteomic Data Commons (PDC), encompassing nine cancer types: stomach adenocarcinoma (STAD), breast invasive carcinoma (BRCA), kidney renal clear cell carcinoma (KIRC), lung adenocarcinoma (LUAD), lung squamous cell carcinoma (LUSC), liver hepatocellular carcinoma (LIHC), head and neck squamous cell carcinoma (HNSC), colon adenocarcinoma (COAD), and uterine corpus endometrial carcinoma (UCEC) (Fig. 3A). These datasets provided complementary representations of transcriptional and protein-level molecular states across tumor and non-malignant tissues. To establish a common interaction framework, we integrated physical PPI evidence from STRING, BioGRID, and IntAct. The initial collection comprised 1,400,832 unique interactions, of which 1,186,437 were retained after applying the broad physical-interaction filtering criteria. Substantial complementarity was observed among the three resources, including 813,454 interactions exclusive to BioGRID, 125,836 exclusive to STRING, and 79,345 exclusive to IntAct, whereas 16,104 interactions were supported by all three databases (Supplementary Fig. S1A). Subsequent gene mapping and quality curation yielded a canonical reference interactome comprising 18,908 proteins and 1,053,426 nonredundant interactions, providing a unified physical scaffold for disease-specific network reconstruction (Supplementary Fig. S1B).

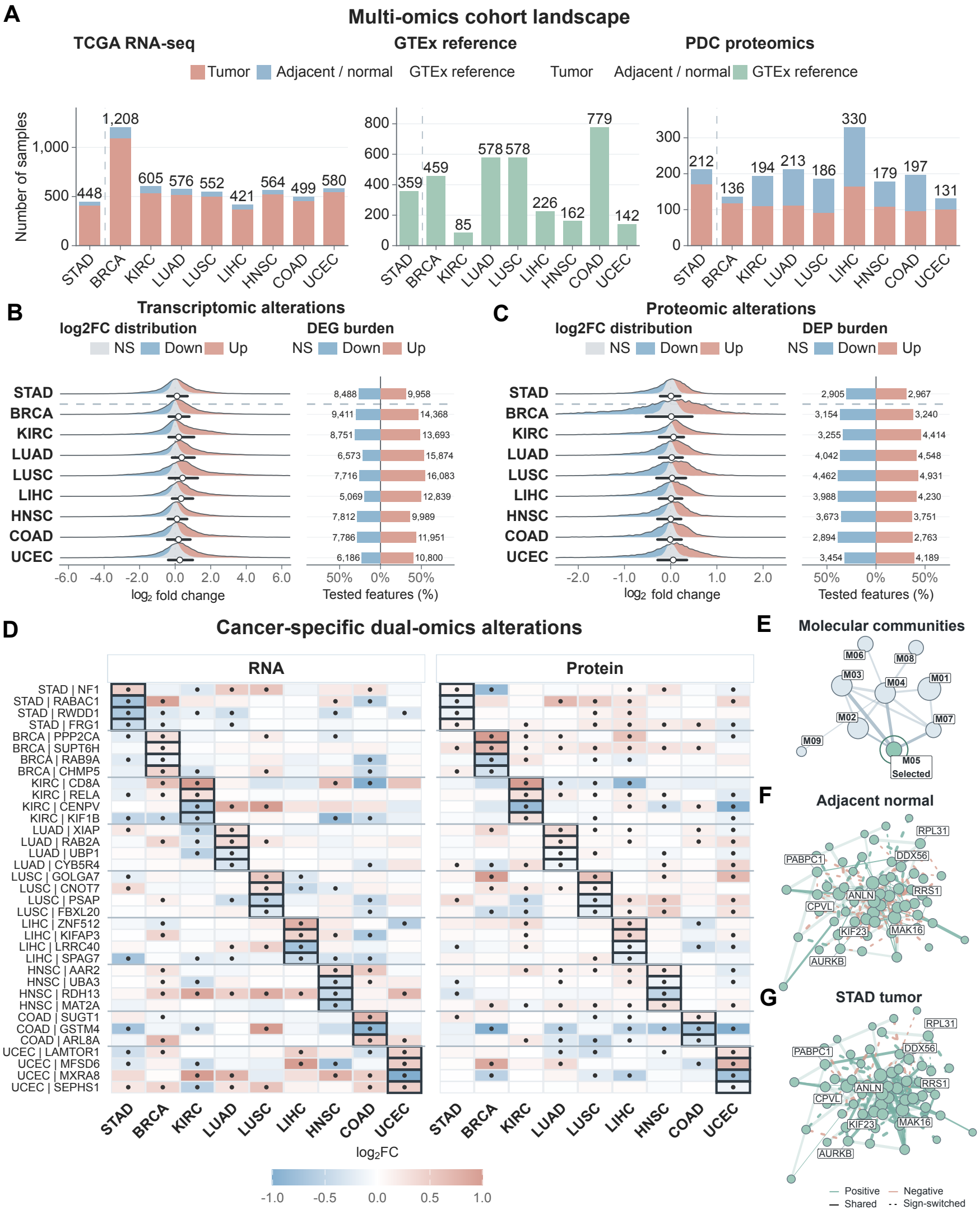

**Fig. 3 Multi-omics characterization and reconstruction of cancer-specific molecular interaction networks. (A)** Overview of transcriptomic and proteomic cohorts collected across nine cancer types. Stacked bar charts show the numbers of tumor and adjacent-normal samples from The Cancer Genome Atlas (TCGA) RNA-sequencing datasets, independent normal-tissue reference samples from the Genotype-Tissue Expression (GTEx) project, and tumor and non-malignant samples from the Proteomic Data Commons (PDC). **(B)** Distribution of tumor-associated transcriptomic alterations across cancer types. Ridgeline plots display the distributions of $\log_2$ fold changes, with negative and positive values representing downregulated and upregulated genes, respectively. Corresponding stacked bars summarize the proportions of significantly downregulated, significantly upregulated and non-significant genes among tested features. **(C)** Corresponding distributions of tumor-associated protein abundance changes and the proportions of differentially abundant proteins (DEPs) across the nine cancers. **(D)** Representative cancer-specific molecular alterations at the transcriptomic and proteomic levels. Heatmaps show $\log_2$ fold changes of selected genes across cancer types, with blue and red indicating decreased and increased abundance, respectively. Outlined blocks identify the corresponding cancer contexts, and symbols indicate statistically significant differential alterations. **(E)** Representative molecular interaction module illustrating the connectivity of proteins within the reference protein–protein interaction network used for multi-omics integration. **(F, G)** Representative stomach adenocarcinoma (STAD)-associated molecular subnetwork reconstructed for adjacent-normal (F) and tumor (G) tissues. The same molecular interaction architecture is shown under the two tissue conditions, illustrating changes in the strengths and signs of molecular associations within a conserved interaction scaffold. Network edges represent

signed, weighted molecular associations inferred through protein-centered integration of proteomic and transcriptomic evidence; positive and negative associations denote concordant and opposing statistical coupling rather than experimentally established biochemical activation or inhibition.

We first characterized cancer-associated molecular alterations at the transcriptomic and proteomic levels (Fig. 3B,C). Differential expression analyses revealed extensive but heterogeneous transcriptional changes across the nine cancer types, accompanied by substantial changes in protein abundance. Although both molecular layers captured prominent tumor-associated alterations, their distributions and relative magnitudes varied considerably among cancers, indicating that transcriptional measurements alone provide an incomplete representation of proteomic changes. Direct comparison of differential expression evidence further demonstrated the complementarity of these modalities (Supplementary Fig. S2A). Among differentially abundant proteins, subsets exhibited concordant transcriptional changes, whereas others showed discordant directions, nonsignificant RNA changes, or unavailable transcriptomic evidence. Conversely, thousands of protein-coding differentially expressed genes in individual cancers lacked corresponding quantitative proteomic measurements. These findings indicate that transcriptomic profiling substantially extends molecular coverage beyond the available proteomic data, whereas proteomics contributes an independent layer of evidence for molecular abundance changes. Because the transcriptomic and proteomic measurements were not uniformly obtained from the same patients, these comparisons characterize cross-cohort molecular evidence rather than within-patient RNA–protein concordance.

To distinguish shared molecular alterations from cancer-restricted expression signatures, we next examined differential expression recurrence across genes statistically evaluable in all nine cancers. Both transcriptomic and proteomic analyses identified extensive recurrence across multiple malignancies, although the numbers of shared differentially expressed molecules progressively decreased as the minimum number of cancer types increased (Supplementary Fig. S2B). Recurrence was more restricted when simultaneous transcriptomic and proteomic evidence was required: 236 genes displayed dual-omics differential expression across all nine cancers, of which 118 additionally satisfied the cross-cancer directional conservation criterion. At the opposite end of the recurrence spectrum, relatively few genes met the stringent definition of significant differential expression in only one of the nine evaluable cancer contexts. Cancer-restricted DEG counts ranged from 2 to 16 per cancer, whereas cancer-restricted DEP counts ranged from 0 to 2 and dual-omics-specific counts from 3 to 12 (Supplementary Fig. S3A). Representative cancer-restricted transcriptomic signatures showed distinct patterns of tumor-associated expression changes across malignancies, with significant alterations concentrated in their corresponding cancer of origin despite variable changes in other cancer types (Supplementary Fig. S3B). Together with the cancer-associated molecular patterns observed in the main analysis (Fig. 3D), these findings reveal that differential expression can be shared across multiple cancer types or restricted to a single cancer under the applied statistical criteria. The limited number of strictly cancer-restricted genes reflects the stringent cross-cancer evaluability and significance criteria rather than an absence of cancer-specific molecular biology.

We next characterized molecular association patterns within tumor and non-malignant tissues, independently of the differential expression analyses. Using the common physical PPI scaffold, we calculated Pearson correlations between PPI-supported molecular pairs separately for each cancer type, tissue state, and omics modality (Supplementary Fig. S2C). The resulting transcriptomic and proteomic correlation distributions were predominantly concentrated around weak associations, with varying positive and negative correlation tails across cancer types. Differences in the distributional profiles were also observed between tumor and non-malignant tissues, suggesting that molecular association patterns may vary across disease states. Unlike differential expression, which characterizes changes in the abundance of individual molecules, correlation-based analysis describes the coordinated variation between molecular pairs. These complementary representations motivated the reconstruction of state-specific, weighted molecular networks to systematically characterize changes in molecular association patterns within a common physical interaction scaffold.

We therefore reconstructed cancer-specific molecular interaction networks by integrating state-dependent molecular associations within the shared physical PPI scaffold. For each cancer, transcriptomic and proteomic profiles were separately used to estimate pairwise molecular correlations among PPI-supported interactions. Network deconvolution was subsequently applied to attenuate indirect association contributions, yielding signed, weighted interaction networks for each molecular layer. To accommodate differences in molecular coverage and measurement stability, we adopted a protein-centered integration strategy in which proteomic associations provided primary interaction evidence, while transcriptomic associations supplied stability-weighted corrections and extended network coverage where sufficient proteomic measurements were unavailable. This integration preserved the common physical interaction prior while allowing individual network edges to exhibit distinct association strengths and signs across pathological states.

We selected STAD as the primary case study to characterize the resulting network representation and demonstrate its suitability for downstream natural-product perturbation modeling. In STAD, 412 tumor and 36 adjacent-normal transcriptomic samples supported an RNA-based network containing 16,809 nodes and 995,313 PPI-restricted interactions, whereas 171 tumor and 41 normal proteomic samples yielded a protein-based network comprising 8,239 nodes and 558,245 interactions. Integration of the two molecular layers generated a connected network containing 16,899 proteins, with 995,799 active interactions in the tumor state and 994,978 in the adjacent-normal state. Across the integrated interaction universe of 998,341 edges, 555,215 interactions (55.6%) were supported by both transcriptomic and proteomic evidence, 440,096 (44.1%) by transcriptomic evidence alone, and 3,030 (0.3%) by proteomic evidence alone (Supplementary Fig. S1C). Thus, proteomic evidence anchored the integrated representation where available, whereas transcriptomic information enabled substantially broader coverage of the cancer-associated molecular system.

Network-level visualization revealed an organized landscape of interconnected molecular communities within the resulting STAD interaction architecture (Fig. 3E). Examination of representative STAD-associated molecular subnetworks further showed that, although numerous interactions were retained across adjacent-normal and tumor states, subsets exhibited changes in their association strengths and signs (Fig. 3F,G). These observations illustrate how molecular relationships can be reorganized within a largely conserved interaction scaffold, capturing changes that are not directly represented by individual RNA or protein abundance measurements. Collectively, the integration of physical interaction evidence, complementary molecular profiling, and state-dependent association inference establishes a disease-contextualized network framework for characterizing cancer molecular organization and provides the foundation for subsequent signed propagation of natural-product perturbations.

### AI-guided target inference and signed network propagation reveal proteome-wide effects of natural products

To investigate how natural products may influence disease-associated molecular systems beyond their immediate protein targets, we established an AI-guided framework integrating large-scale compound–protein interaction prediction, regulatory-direction inference, and signed network propagation. Using the standardized natural-product registry, we performed proteome-wide interaction inference with three complementary models, including ConPLex, DTIAM-Merged, and DTIAM-Yamanishi, against a unified panel of 19,281 human proteins. The individual models exhibited distinct prediction-score distributions, with substantial differences in the score ranges occupied by their highest-ranking compound–protein associations (Supplementary Fig. S4A). To integrate these heterogeneous outputs, we employed a weighted consensus strategy that combined model-specific interaction evidence while accounting for differences in their predictive representations and scoring scales. This approach identified 17,772,830 candidate compound–protein pairs with weighted DTI support of at least 0.4. Increasing the support threshold to 0.6, 0.8, and 1.0 progressively reduced the candidate set to 828,477, 145,676, and 29,280 pairs, respectively (Fig. 4A). Thus, the consensus framework generated a broad predicted interaction landscape while distinguishing a substantially smaller subset of associations with stronger combined computational support.

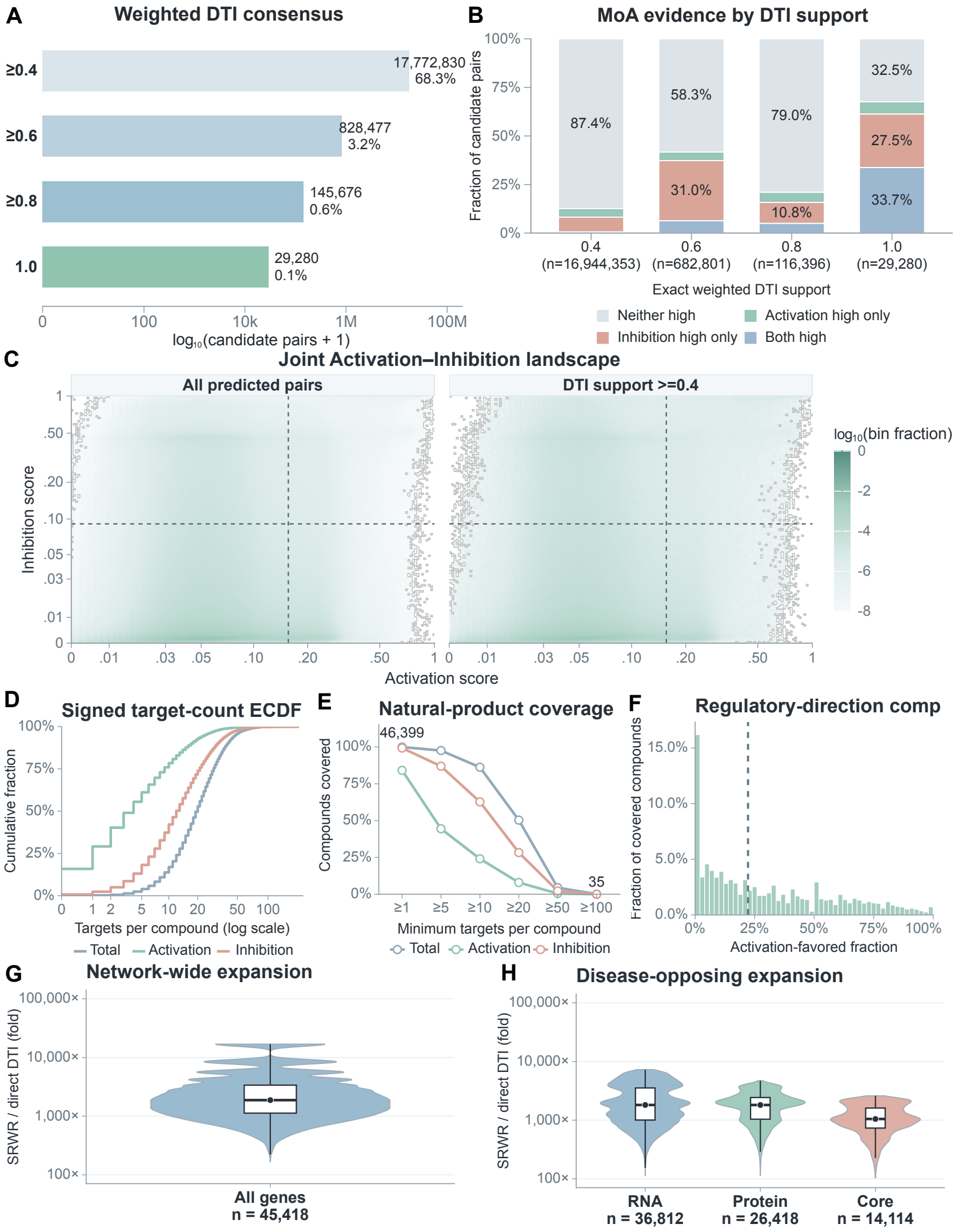

**Fig. 4 AI-guided compound–target inference and signed network propagation enable genome-scale molecular influence modeling of natural products. (A)** Distribution of candidate compound–protein interactions across cumulative weighted drug–target interaction (DTI) support thresholds. Consensus scores were derived from three complementary AI models, and horizontal bars indicate the number and proportion of compound–protein pairs retained at each threshold. **(B)** Distribution of inferred regulatory-direction evidence across exact weighted DTI

support strata. Compound–protein pairs were classified as activation-high only, inhibition-high only, both high or neither high, using the 95th-percentile thresholds of the respective activation and inhibition score distributions. Stacked bars indicate the proportion of each category within individual DTI support strata. **(C)** Joint distributions of activation- and inhibition-associated scores across all inferred compound–protein pairs (left) and pairs satisfying weighted DTI support ≥0.4 (right). Color intensity represents the logarithmically scaled fraction of compound–protein pairs within each two-dimensional score bin, and dashed lines indicate reference score thresholds. **(D)** Empirical cumulative distribution functions (ECDFs) of activation-associated, inhibition-associated and total direction-resolved target counts per natural compound. **(E)** Natural-product coverage as a function of the minimum number of retained targets per compound, shown separately for total, activation-associated and inhibition-associated target sets. **(F)** Distribution of the activation-favored fraction among direction-resolved targets across natural compounds, illustrating heterogeneity in inferred regulatory-direction composition. **(G)** Distribution of network-wide molecular influence expansion following signed random walks with restart (SRWR) in the stomach adenocarcinoma (STAD)-specific molecular network. Expansion was calculated as the number of network nodes receiving nonzero propagated influence scores relative to the number of mapped direct signed targets for each compound (n = 45,418). **(H)** Distributions of disease-opposing molecular influence expansion across transcriptomic (RNA; n = 36,812), proteomic (Protein; n = 26,418) and concordant RNA–protein (Core; n = 14,114) disease signatures. Expansion represents the ratio of disease-associated molecules receiving oppositely directed network influence scores to those covered by oppositely directed direct-target seeds. Violin plots show the distributions of compound-level expansion ratios, with embedded boxplots summarizing their central tendency and dispersion. The distributions in G and H quantify the expansion of computationally inferred molecular influence rather than experimentally validated protein responses.

Because predicted compound–target interactions do not inherently indicate whether a compound activates or inhibits its associated proteins, we next incorporated independently inferred activation and inhibition scores to characterize the potential regulatory direction of candidate interactions. The two score distributions exhibited markedly different characteristics, with activation scores occupying a broader range and inhibition scores concentrated predominantly near zero (Supplementary Fig. S4B). Stratification by exact weighted DTI support revealed substantial heterogeneity in regulatory-direction evidence (Fig. 4B). Among candidate pairs in the lowest retained support stratum (0.4), 87.4% exhibited neither high activation nor high inhibition evidence. In contrast, among pairs in the highest support stratum (1.0), this fraction decreased to 32.5%, while 33.7% exhibited simultaneously high activation and inhibition scores and 27.5% exhibited inhibition-high-only evidence. The coexistence of strong interaction support and ambiguous regulatory-direction predictions indicates that interaction confidence and inferred mode of action represent distinct dimensions of compound–target evidence. In particular, a highly supported interaction cannot automatically be assigned an activating or inhibitory effect without additional directional information.

We further characterized the joint distribution of predicted activation and inhibition scores across the complete compound–protein prediction space and the subset satisfying the minimum weighted DTI support threshold (Fig. 4C). The score landscapes were heterogeneous, containing regions with comparatively low evidence for both regulatory directions, asymmetric activation- or inhibition-associated evidence, and simultaneously elevated scores. Comparison of the DTI-supported subset with the broader prediction background revealed localized enrichment within the activation–inhibition score space (Supplementary Fig. S4C), indicating that interaction-based selection altered the representation of different predicted regulatory states. These findings motivated the use of explicit directional-evidence criteria to construct signed compound–target associations, rather than imposing a regulatory sign solely on the basis of predicted interaction support. Importantly, the resulting activation- and inhibition-associated categories describe computationally inferred regulatory tendencies, not experimentally established agonistic or antagonistic activities.

We next investigated how these predicted directional relationships were distributed across the natural-product collection (Fig. 4D–F). The number of direction-resolved targets varied substantially among compounds, with distinct cumulative distributions for activation-associated, inhibition-associated, and total signed target counts (Fig. 4D). Activation-associated targets were generally less numerous, whereas inhibition-associated and combined target profiles extended toward larger predicted target sets. A total of 46,399 compounds were represented in the direction-resolved target landscape, but compound coverage declined progressively as the minimum required target count increased (Fig. 4E). At a requirement of at least 100 targets per compound, only 35 compounds remained represented. The relative contributions of activation- and inhibition-favored associations were also heterogeneous across compounds (Fig. 4F), revealing a wide spectrum of predicted regulatory-direction compositions rather than a uniform balance of activating and inhibitory effects. We additionally examined the sensitivity of natural-product and target-gene coverage to increasingly stringent raw and percentile-based activation–inhibition differences (Supplementary Fig. S4F). Natural-product coverage remained nearly complete across the evaluated thresholds, whereas the number of retained target genes was more sensitive to directional filtering. At the most stringent displayed raw-score difference of 0.20, 13,720 human target genes (71.2%) remained covered; under a percentile-based difference of 0.30, coverage was 16,351 genes (84.8%). These results demonstrate that directional-evidence filtering can substantially reshape the molecular breadth of inferred compound–target profiles while retaining most natural products within the analysis.

To translate these direct compound–target predictions into disease-contextualized molecular influence profiles, we integrated the signed interaction evidence with the STAD-specific multi-omics network. Candidate direct targets were assigned signed perturbation weights based on their combined interaction support and relative activation–inhibition evidence. These weights served as initial seeds for SRWR on the protein-centered STAD tumor network, comprising 16,899 molecular nodes. The number of signed targets successfully mapped to this network differed considerably among compounds, producing a broad distribution of initial seed availability (Supplementary Fig. S4E). SRWR then propagated the initial positive and negative signals through the weighted, signed molecular interaction network, accounting for network connectivity and the strengths and directions of molecular associations. Unlike direct target inference, which provides information only for selected compound–protein pairs, this procedure assigns continuous directional influence scores to additional network-connected molecules. The resulting profiles therefore represent inferred network-mediated influence within the STAD disease context, rather than direct evidence of compound binding or measured changes in molecular abundance.

Across 45,418 compounds with valid signed targets mapped to the STAD network, SRWR substantially extended molecular coverage beyond the original direct-target predictions (Fig. 4G). For compounds with defined expansion ratios, the number of network-influenced genes was typically several orders of magnitude greater than the number of initial signed targets, reflecting the capacity of propagation to distribute inferred perturbation signals across the connected molecular system. To determine how directly predicted targets contributed to the resulting rankings, we compared original signed seeds with network-prioritized nonseed genes at successive SRWR ranking depths (Supplementary Fig. S4D). Original signed targets accounted for approximately 72% of the top 10 ranked compound–gene pairs, but their contribution progressively decreased with increasing ranking depth, reaching approximately 5.5% among the top 200. Accordingly, broader SRWR rankings were dominated by genes prioritized through network connectivity rather than direct signed compound–target evidence. These observations demonstrate that network propagation changes not only the number of molecular associations represented for each compound, but also the composition of its prioritized molecular influence profile.

We subsequently assessed whether this expanded representation could capture potential molecular perturbations opposing the STAD disease state. Three complementary disease signatures were considered: transcriptomically defined differentially expressed genes, proteomically defined differentially abundant proteins, and a concordant RNA–protein core. For each signature, we evaluated the expansion from directly predicted signed targets to network-influenced molecules whose inferred perturbation directions opposed their tumor-associated alterations (Fig. 4H). SRWR produced substantial disease-opposing expansion across all three molecular layers, with defined positive expansion ratios available for 36,812 compounds in the RNA signature, 26,418 in the protein signature, and 14,114 in the concordant core. The distributions indicated approximately three orders of magnitude of expansion for many eligible compounds, although the magnitude varied across molecular signatures and individual compounds. The distinct evaluation populations reflect differences in the availability of direct disease-opposing targets and propagated scores; they should not be interpreted as evidence that every screened compound exhibits the same magnitude of disease-opposing activity. Nevertheless, these findings establish that signed propagation can extend potential disease-opposing influence beyond directly predicted targets across complementary transcriptomic and proteomic representations.

Collectively, these analyses establish a scalable framework that converts heterogeneous AI-inferred compound–protein associations into signed, disease-contextualized molecular influence profiles. By separating predicted interaction support from regulatory-direction evidence and incorporating the topology of a multi-omics-informed cancer network, the framework extends candidate molecular perturbations beyond direct compound targets while preserving their inferred directional relationships. Although network expansion alone does not establish therapeutic activity, the resulting profiles provide a systematic basis for assessing whether natural-product-associated molecular influences preferentially oppose disease-associated alterations and for evaluating these predictions against independent biological evidence.

**Orthogonal molecular evidence supports disease-opposing natural-product effects and herb-level modeling**

To evaluate the biological relevance of the inferred network influence profiles, we investigated whether natural products associated with gastric cancer exhibited preferentially disease-opposing molecular effects within the STAD-specific network. We curated a literature-supported collection of 35 gastric cancer-associated compounds, of which 34 possessed eligible signed targets and corresponding SRWR profiles. These compounds were evaluated against 10,000 sets of complexity-matched random compounds, each containing 34 controls matched for direct-target characteristics. Three complementary disease signatures were considered: 10,262 differentially expressed genes, 5,783 differentially abundant proteins, and a concordant RNA–protein core comprising 3,064 genes with significant and directionally consistent alterations at both molecular levels. For each signature, disease-opposing influence was defined as a negative predicted effect on tumor-upregulated molecules or a positive predicted effect on tumor-downregulated molecules. This design enabled us to assess whether disease-associated compounds preferentially exhibited molecular influence patterns opposing the STAD disease state rather than merely producing widespread network perturbations.

We first examined directional recovery within the concordant RNA–protein core, prioritizing genes exhibiting the strongest combined disease-associated alterations (Fig. 5A). Among the top 10% most strongly dysregulated core genes, literature-supported compounds exhibited higher predicted suppression of disease-upregulated molecules than matched controls (50.3% versus 44.6%; empirical P = 0.0302). A more pronounced difference was observed for restoration of disease-downregulated molecules (48.4% versus 36.1%; P = 0.0048), resulting in greater combined directional recovery (46.3% versus 40.2%; P = 0.0415). Similar patterns were observed when the analysis was extended to the top 20% of concordantly dysregulated genes, with predicted disease-up suppression, disease-down restoration, and combined recovery of 47.0%, 44.7%, and 47.1%, respectively, compared with 40.7%, 37.1%, and 38.9% for matched controls (P = 0.0479, 0.0423, and 0.0173, respectively). Notably, the top 10% and top 5% disease-down restoration metrics remained significant after correction within the six predefined disease-strength thresholds of the DOWN-recovery family (adjusted q = 0.0288 and 0.0342, respectively), although neither survived correction across all 18 concordant-core recovery comparisons. These findings indicate that the strongest relative disease-opposing signals were concentrated among molecules jointly downregulated at the RNA and protein levels.

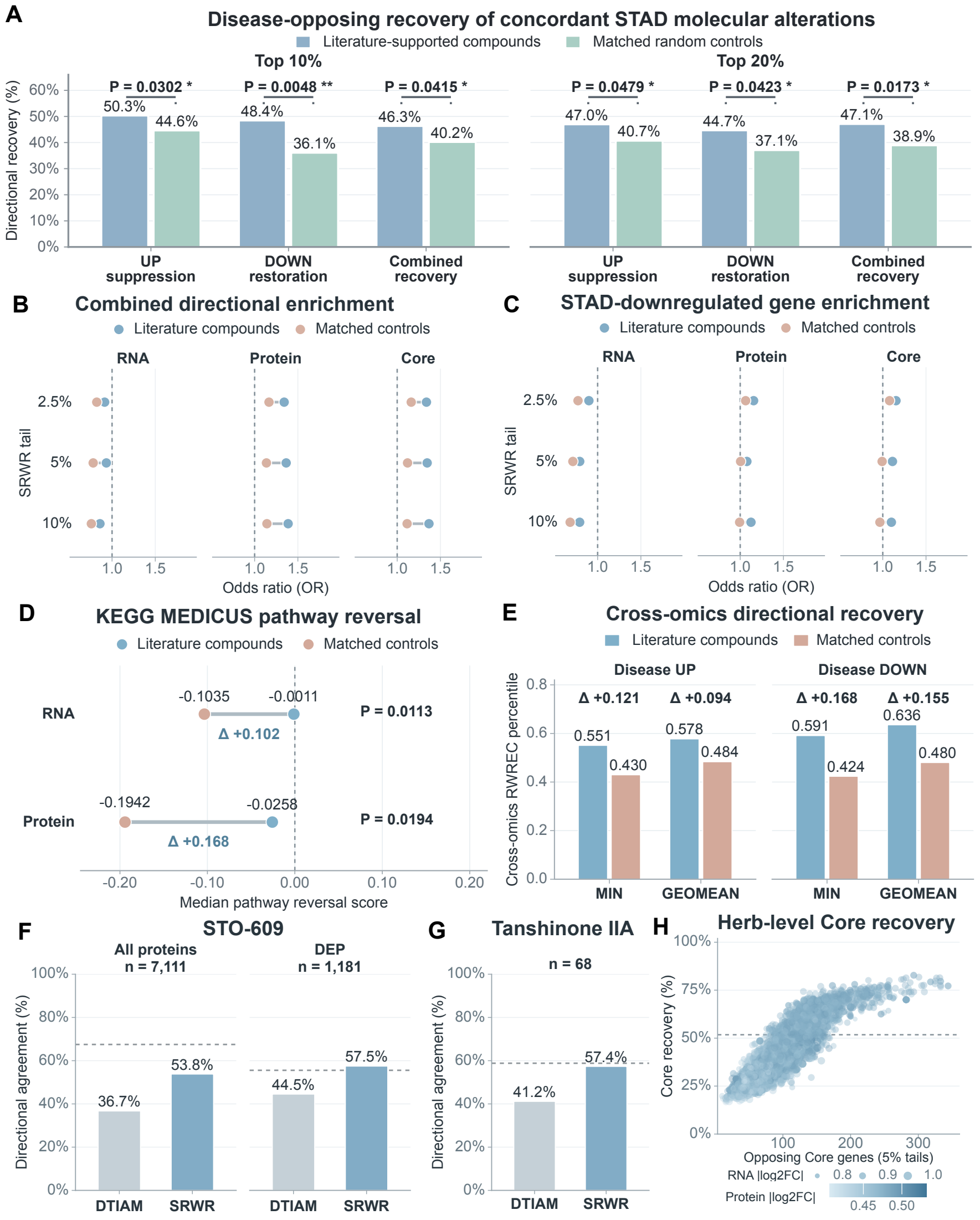

**Fig. 5 Evaluation of disease-opposing molecular influences and herb-level modeling in gastric cancer. (A)** Comparison of predicted directional recovery between 34 literature-supported gastric cancer compounds and complexity-matched random controls within the concordant RNA–protein disease signature. Recovery was evaluated among the top 10% and 20% of disease-associated genes ranked by molecular alteration strength, separately for suppression of disease-upregulated genes, restoration of disease-downregulated genes and their combined recovery. Bars represent median compound-level recovery rates, and annotated P values indicate empirical comparisons against 10,000 matched random sets. **(B, C)** Enrichment of strongly inferred disease-opposing network influences among transcriptomic (RNA), proteomic (Protein) and concordant RNA–protein (Core) signatures. Odds ratios quantify the enrichment of disease-associated genes within the strongest positive and negative signed random walk with restart (SRWR) influence tails (2.5%, 5% and 10%). B shows combined directional enrichment, whereas C focuses on positive network influences associated with restoration of STAD-downregulated genes. Points compare literature-supported compounds with matched controls; dashed vertical lines indicate an odds ratio of 1. **(D)** Directional pathway reversal analysis using KEGG MEDICUS gene sets. Median pathway reversal scores are shown for RNA- and protein-derived disease signatures, comparing literature-supported compounds with matched controls. Positive score differences indicate a relative shift toward disease-opposing molecular influences. **(E)** Cross-omics concordance of directional recovery, integrating RNA- and protein-derived rank-weighted recovery scores through minimum (MIN) and geometric-mean (GEOMEAN) operators. Bars show median percentile-based consensus scores for disease-upregulated and disease-downregulated signatures, with annotated differences indicating the improvement relative to matched controls. **(F, G)** Retrospective evaluation of inferred molecular influence directions against experimentally measured drug-induced protein abundance changes. F compares direct DTIAM-derived regulatory-direction scores with STAD-specific SRWR for STO-609 across all evaluable proteins (n = 7,111) and the differentially abundant protein subset (n = 1,181). G shows the corresponding comparison for the natural compound tanshinone IIA using 68 proteins with consistent experimental directional annotations. Bars indicate the proportion of proteins for which the inferred direction agrees with the experimentally observed abundance change. **(H)** Exploratory extension of molecular influence modeling to multicomponent medicinal herbs. Scatter plot showing the relationship between the number of concordant STAD disease genes receiving oppositely directed influences within the strongest 5% SRWR tails and herb-level Core recovery across the indicated RNA and protein disease-effect thresholds. The horizontal dashed line denotes 50% recovery. Herb-level profiles represent computationally inferred molecular influences rather than experimentally established therapeutic effects.

Comparative compound-level analyses used the same frozen set of 10,000 complexity-matched control groups, each containing 34 compounds. Reported empirical P values should be interpreted within the exploratory evaluation framework; nominal or within-family significance does not necessarily imply significance after correction across all evaluated endpoints.

We next assessed whether the most strongly influenced network nodes preferentially overlapped with disease-associated molecular signatures (Fig. 5B,C). Directional extreme-tail enrichment was quantified by comparing the strongest negative and positive SRWR scores with the corresponding disease-upregulated and disease-downregulated gene sets. Across RNA, protein, and concordant-core signatures, the analysis identified heterogeneous enrichment patterns, with stronger relative separation for selected disease-associated molecular subsets. In particular, enrichment of the most strongly negative network effects among disease-upregulated genes and positive network effects among disease-downregulated genes provided complementary views of the specificity of predicted network influence. However, the aggregate extreme-tail analyses did not consistently withstand correction for multiple comparisons, indicating that the observed advantages were dependent on the molecular signature and the selected influence-strength range. These results indicate that the relative enrichment of predicted disease-opposing effects varies across molecular signatures and SRWR influence thresholds, with no consistent advantage over matched controls after correction for multiple comparisons.

To determine whether the predicted effects extended to coordinated biological programs, we evaluated directional pathway reversal using curated pathway collections, including KEGG MEDICUS, Reactome, Hallmark, and Gene Ontology Biological Process (Fig. 5D). Pathway scores integrated the direction and relative strength of disease-associated molecular alterations with rank-normalized compound SRWR profiles. Among the evaluated pathway collections, KEGG MEDICUS provided the clearest aggregate separation between literature-supported compounds and matched controls. The median RNA-based pathway reversal score shifted from −0.1035 in matched controls to −0.0011 in literature-supported compounds ($P = 0.0113$), whereas the corresponding protein-based score increased from −0.1942 to −0.0258 ($P = 0.0194$). Both comparisons remained significant after correction within the six KEGG aggregate tests ($q = 0.0339$ and 0.0388, respectively), but not after correction across the broader pathway-database comparison family. Although these shifts did not establish absolute reversal of the entire pathway landscape, they indicated that literature-supported compounds were less aligned with STAD-associated molecular alterations than complexity-matched controls.

We further investigated whether individual compounds exhibited concordant disease-opposing influence at both the transcriptomic and proteomic levels (Fig. 5E). RNA- and protein-based recovery scores were independently converted to percentile ranks and combined using either a conservative minimum operator or their geometric mean. For disease-upregulated molecular programs, literature-supported compounds showed modest relative improvements, with minimum-based and geometric-mean percentile scores of 0.551 and 0.578, compared with 0.430 and 0.484 for matched controls. More pronounced differences emerged for disease-downregulated programs, for which the corresponding scores reached 0.591 and 0.636 in literature-supported compounds, compared with 0.424 and 0.480 in matched controls ($P =$ 0.0182 and 0.0198, respectively). These comparisons showed nominal evidence of coordinated disease-down restoration across both molecular layers, although neither survived correction across the six cross-omics comparisons. Together with the concordant-core analysis, these findings support a preferential, but not uniformly significant, tendency toward predicted restoration of molecular programs suppressed in STAD.

To obtain an orthogonal assessment using experimentally measured drug-induced molecular responses, we compared predicted regulatory directions with independent treatment-associated proteomic changes for STO-609 and tanshinone IIA (Fig. 5F,G). The pharmacological tool compound STO-609 served as a methodological reference, whereas tanshinone IIA represented a natural-product case with published gastric cancer cell perturbation data. For STO-609, the directional agreement between direct activation–inhibition predictions and experimental protein abundance changes was 36.7% across 7,111 evaluable proteins, increasing to 53.8% with STAD-specific SRWR. Among 1,181 differentially abundant proteins, the corresponding agreement increased from 44.5% to 57.5%. For tanshinone IIA, analysis of 68 proteins with consistent experimental directional annotations showed an increase in agreement from 41.2% for direct predictions to 57.4% following network propagation. Although the tanshinone IIA improvement did not reach statistical significance in the paired comparison (exact McNemar $P = 0.0989$), these observations suggest that incorporating disease-associated network connectivity can improve alignment with experimentally observed protein abundance directions in selected perturbation contexts. The moderate overall agreement also highlights the distinction between inferred network influence and experimentally measured protein responses.

Finally, we extended the analytical framework from individual natural products to multicomponent herbal systems by connecting standardized herb compositions with compound-level network influence profiles (Fig. 5H). Herb-level predictions were examined against the concordant STAD molecular core, focusing on disease-opposing signals among strongly influenced network genes. The resulting analysis revealed heterogeneous predicted recovery across herbal compositions and disease-effect thresholds, illustrating how the standardized herb–compound atlas could be linked to the same disease-specific molecular framework used for individual compounds. This extension provides an exploratory systems-level representation of herbal interventions, although the inferred effects do not by themselves establish experimentally validated herb activity or synergistic interactions among constituents. Collectively, these analyses provide complementary evidence for the biological relevance of disease-contextualized network influence modeling and demonstrate its potential extension from individual natural compounds to compositionally complex herbal systems.

**Pan-cancer network rewiring reveals disease-specific molecular alterations beyond differential expression**

To investigate whether the multi-omics-informed network framework could reveal molecular alterations beyond the STAD case study, we extended our analyses to nine cancer types and systematically compared their molecular network architectures and disease-associated interaction changes. As expected from the shared physical PPI reference, the reconstructed cancer-specific networks exhibited highly overlapping interaction architectures (Fig. 6A,B). These networks contained 16,095–17,089 molecular nodes and approximately 0.966–1.001 million active interactions, with mean node degrees ranging from 117.2 to 120.0. Across cancer types, 90.1–95.7% of nodes and 91.9–95.3% of edges were shared, reflecting the common interaction scaffold underlying network reconstruction. Importantly, this architectural similarity does not necessarily imply conservation of disease-associated molecular relationships. Rather, the shared interaction scaffold provides a consistent structural basis for investigating cancer-specific alterations in molecular association strengths and interaction patterns, allowing disease-associated network rewiring to be characterized independently of broad differences in network topology.

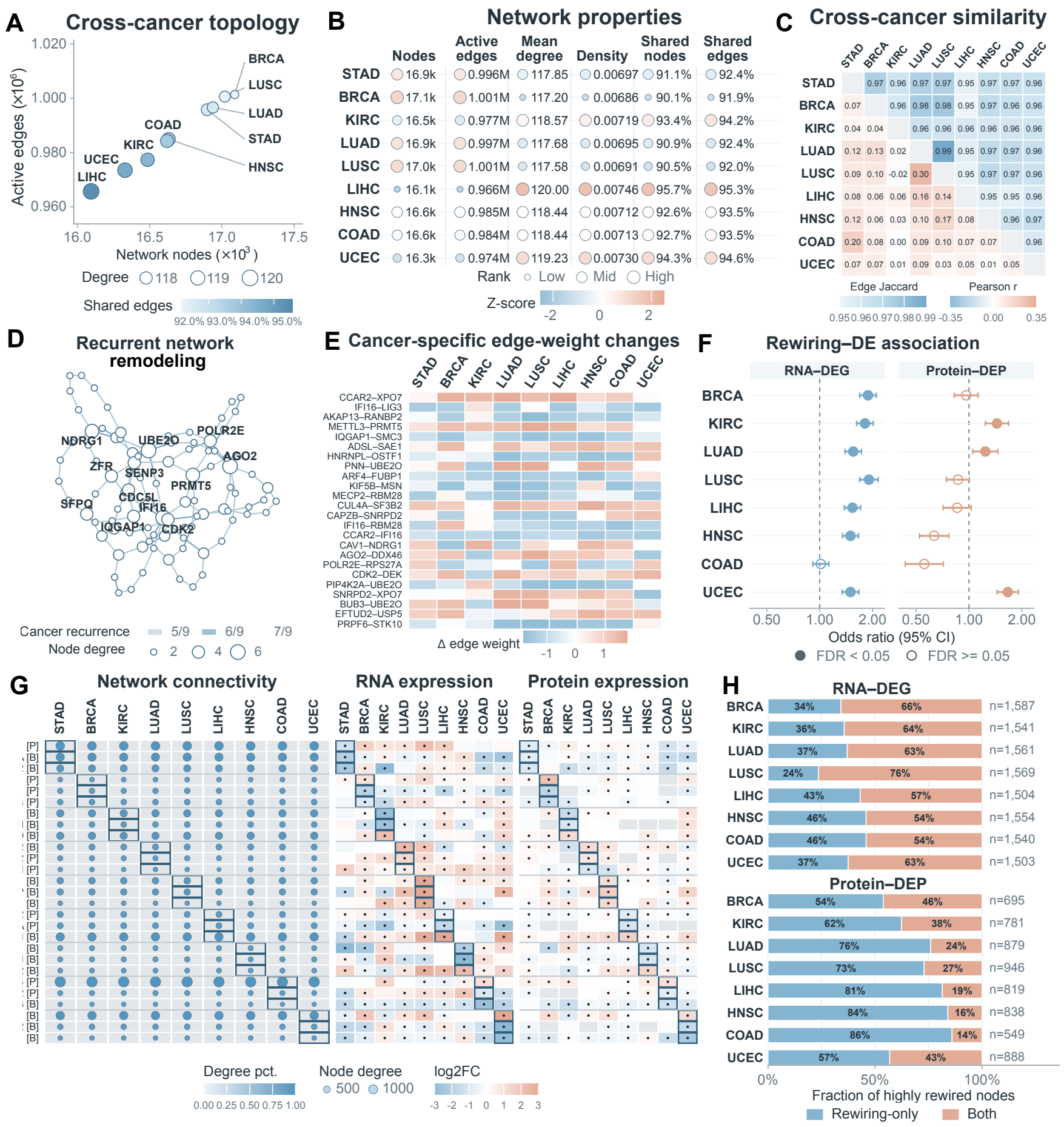

**Fig. 6 Pan-cancer molecular networks reveal disease-specific rewiring within conserved interaction architectures. (A)** Comparison of network size and connectivity across nine cancer-specific molecular networks. Scatter plot shows the relationship between the number of molecular nodes and active interactions, with point size representing mean node degree and color indicating the proportion of shared interactions. **(B)** Summary of network topological properties across the nine cancer types, including node and active-edge counts, mean degree, network density, and the proportions of shared nodes and edges. Symbol sizes and colors represent the relative magnitude of each property, with color encoding based on column-wise standardized values. **(C)** Pairwise comparison of molecular network similarity across cancer types. Heatmaps display Jaccard similarity of interaction sets and Pearson correlations of disease-associated changes in signed interaction weights, distinguishing conserved network architecture from heterogeneous molecular rewiring. **(D)** Representative connected subnetwork comprising proteins recurrently affected by substantial molecular interaction remodeling across multiple cancers. Node characteristics represent network connectivity and the recurrence of remodeling across cancer types. **(E)** Heatmap showing cancer-associated changes in signed interaction weights for representative molecular pairs across the nine cancer types. Blue and red indicate negative and positive changes in interaction weights, respectively, illustrating heterogeneous patterns of molecular association remodeling. **(F)** Association between differential molecular abundance and network rewiring across eight additional cancer types. Forest plots show odds ratios (ORs) and 95% confidence intervals for the enrichment of genes with pronounced differential expression among highly rewired nodes, evaluated separately for transcriptomic (RNA–DEG) and proteomic (Protein–DEP) networks. Filled points indicate associations significant after Benjamini–Hochberg false discovery rate correction (FDR < 0.05), whereas open points indicate non-significant associations. Dashed vertical lines denote OR = 1. **(G)** Comparison of network connectivity and differential molecular abundance for representative genes across the nine cancer types. Dot plots display relative

node connectivity, whereas adjacent heatmaps show tumor-associated RNA and protein $\log_2$ fold changes. Blue and red indicate negative and positive expression changes, respectively, highlighting the heterogeneous correspondence between network organization and molecular abundance. **(H)** Proportions of highly rewired nodes with or without significant differential molecular abundance across the eight additional cancer types. Stacked bars distinguish rewiring-only nodes from those exhibiting both network rewiring and differential expression, separately for transcriptomic and proteomic networks. Percentages represent the corresponding fractions among highly rewired nodes, and sample sizes indicate the number of evaluated nodes. These analyses demonstrate that network remodeling captures molecular alterations not fully represented by conventional differential expression.

We next examined whether this structural similarity translated into comparable patterns of disease-associated molecular network remodeling. Pairwise comparisons revealed consistently high edge-set similarity across the nine cancers, with Jaccard indices ranging from approximately 0.95 to 0.99 (Fig. 6C). In striking contrast, the correlations between cancer-associated edge-weight changes were substantially lower, with Pearson correlation coefficients ranging from approximately −0.02 to 0.30. Even among cancers with nearly identical interaction coverage, the direction and magnitude of molecular association changes differed considerably. The greatest similarity was observed between lung adenocarcinoma and lung squamous cell carcinoma, whereas most other cancer pairs exhibited substantially weaker correspondence in their interaction changes. These findings reveal a fundamental distinction between the conservation of molecular interaction architecture and the disease-specific reorganization of interaction strengths, suggesting that shared molecular connectivity can support diverse pathological network states.

To identify molecular features underlying these shared and cancer-specific patterns, we investigated recurrent network remodeling across the nine malignancies (Fig. 6D). This analysis revealed a connected subnetwork containing proteins repeatedly associated with substantial interaction changes in five to seven cancer types, including PRMT5, AGO2, IFI16, IQGAP1, CDK2, UBE2O, and POLR2E. Although these proteins participated in recurrently remodeled network regions, their individual interaction changes were not necessarily conserved in direction across cancers. Examination of representative protein pairs further revealed pronounced cancer-dependent differences in signed interaction weights (Fig. 6E). Interactions such as CCAR2–XPO7, METTL3–PRMT5, IQGAP1–SMC3, and AGO2–DDX46 exhibited heterogeneous strengthening, weakening, or changes in association direction across cancer types. Together, these observations indicate that recurrently affected molecular components can undergo distinct patterns of interaction reorganization depending on the pathological context, highlighting the coexistence of shared network vulnerability and cancer-specific molecular remodeling.

We subsequently investigated the relationship between network rewiring and conventional differential expression to determine whether interaction-level changes reflected broader disease-associated molecular abnormalities (Fig. 6F). Across the eight additional cancers, genes displaying pronounced transcriptional alterations were significantly enriched among highly rewired RNA-network nodes in seven cancer types, with odds ratios ranging from approximately 1.50 to 1.91 after multiple-testing correction. COAD represented an exception, showing no significant enrichment. At the proteomic level, the association was less consistent: significant enrichment of highly rewired nodes among proteins with pronounced differential abundance was observed in KIRC, LUAD, and UCEC, whereas the remaining cancers did not exhibit significant positive associations. These results demonstrate that network remodeling is related to differential molecular abundance, but the strength and consistency of this relationship vary considerably between transcriptomic and proteomic layers and across cancer types.

To further characterize the relationship between molecular abundance and network organization, we examined representative proteins across all nine cancers, jointly comparing their network connectivity with tumor-associated changes in RNA and protein abundance (Fig. 6G). The selected proteins exhibited heterogeneous connectivity patterns across cancer types, even when their RNA or protein abundance changes were relatively modest. Conversely, marked transcriptional or proteomic alterations were not uniformly accompanied by corresponding differences in network connectivity. These observations illustrate the imperfect correspondence between molecular abundance and network organization, suggesting that changes in molecular interactions and changes in individual molecular quantities capture related but distinct aspects of cancer-associated molecular states.

We finally quantified the extent to which highly rewired molecular nodes could be identified through conventional differential expression analysis alone (Fig. 6H). Across the eight additional cancer types, 24–46% of highly rewired RNA-network nodes did not meet the corresponding differential gene expression criteria. This discrepancy was more pronounced at the proteomic level, where 54–86% of highly rewired protein-network nodes were not classified as differentially abundant proteins. Notably, the proportion of rewiring-only protein nodes reached 81% in LIHC, 84% in HNSC, and 86% in COAD. Thus, although differential expression and network remodeling overlap for subsets of cancer-associated molecules, a substantial fraction of highly remodeled molecular interactions involves proteins without statistically significant abundance changes. These findings suggest that conventional differential expression analyses alone may overlook potentially informative alterations in the organization of cancer-associated molecular networks.

Collectively, our pan-cancer analyses reveal that largely conserved molecular interaction architectures undergo heterogeneous, cancer-specific reorganization, with network remodeling capturing molecular features that are not fully represented by transcriptional or proteomic differential expression. Beyond providing a disease-contextualized substrate for natural-product perturbation modeling, multi-omics-informed molecular networks therefore offer a complementary systems-level representation of pathological molecular states. By linking changes in molecular abundance with alterations in interaction organization, this framework provides an avenue for investigating both shared and cancer-specific molecular processes and for contextualizing potential therapeutic interventions within disease-associated biological networks.

## Discussion

In this study, we established an integrative computational framework for systematically characterizing the potential proteome-wide molecular effects of natural compounds and multicomponent medicinal herbs within disease-specific biological contexts. By harmonizing multiple herbal medicine databases, we constructed a standardized knowledge atlas encompassing 46,422 chemical entities, 12,459 medicinal herbs, and 156,588 herb–compound associations. Integrating proteomic and transcriptomic data with high-confidence protein–protein interactions further enabled the reconstruction of disease-specific molecular networks across nine cancer types, revealing extensive molecular rewiring within largely conserved interaction architectures. Building on these resources, we combined AI-based compound–target inference with SRWR to extend sparse predicted direct interactions into network-wide, direction-resolved molecular influence profiles. Analyses of gastric cancer demonstrated the feasibility of this approach for prioritizing disease-opposing molecular influences, while retrospective perturbation evidence and exploratory herb-level analyses provided additional assessments of its biological relevance. Together, these findings establish a unified computational framework connecting natural-product chemistry, disease-specific molecular organization, and systems-level therapeutic hypotheses.

Modeling the genome-scale molecular consequences of natural compounds, and particularly multicomponent medicinal herbs, remains a fundamental challenge in natural-product drug discovery[14,28]. Although large-scale perturbation resources and recent deep-learning models have enabled increasingly comprehensive prediction of compound-induced transcriptional responses, these approaches primarily characterize changes in gene expression rather than protein abundance or protein-associated molecular functions[16,17,29]. This distinction is particularly important because transcript abundance does not consistently reflect protein abundance, and molecular responses may involve changes in protein interactions and biological pathways that are not captured by differential expression alone[20,21]. Moreover, direct compound–target prediction provides an inherently sparse representation of potential molecular engagement, whereas experimentally characterized target interactions and proteome-wide perturbation responses remain insufficiently characterized for many natural compounds. Our framework addresses these limitations by anchoring disease-specific molecular networks in available proteomic evidence, using transcriptomic associations to extend molecular coverage, and propagating predicted directional target signals across the resulting networks. This strategy builds on the principle that drug-associated molecular influences may extend beyond directly engaged targets through interconnected biological systems[27]. Importantly, our pan-cancer analyses revealed that disease-associated network rewiring was not fully explained by differential gene or protein abundance, emphasizing the complementary information provided by molecular interaction patterns. Rather than treating network propagation as a direct substitute for experimental perturbation measurements, our approach provides a scalable means of generating protein-referenced, disease-contextualized molecular influence hypotheses for natural compounds and extending their analysis to complex herbal compositions.

Beyond conventional phytochemicals, the broader applicability of this framework may extend to other biologically active constituents of medicinal herbs. Emerging studies have implicated plant-derived regulatory RNAs in antiviral activity and microbiota-mediated host responses[30,31]. Likewise, amino acids can function as signaling molecules through specific protein-sensing mechanisms, as demonstrated by leucine- and arginine-dependent regulation of mTORC1[32,33]. These findings highlight the diversity of molecular mechanisms through which herbal constituents may influence biological systems, extending beyond the conventional pharmacological characterization of secondary metabolites. Notably, the modular architecture of our framework separates the inference of initial molecular interactions from the downstream propagation of their potential effects through disease-specific networks. This design offers a conceptual basis for incorporating additional molecular classes through appropriate interaction and regulatory evidence, including RNA–transcript targeting relationships and amino acid–protein sensing mechanisms. Such extensions could enable the construction of genome-scale, protein-network influence profiles for a broader range of herbal constituents, providing an opportunity to investigate their potential contributions to the system-level actions of medicinal herbs. Importantly, these applications would require molecular-class-specific validation and, where relevant, additional representations of indirect mechanisms such as microbiota-mediated host regulation.

Several limitations warrant consideration. First, the present study remains predominantly computational, without newly performed prospective experiments to establish direct target engagement, protein-level perturbation responses, or therapeutic efficacy. Although retrospective comparisons with published drug-treatment proteomic data provided preliminary support for selected predictions, these analyses were limited in scope and cannot establish general predictive accuracy. Second, the incomplete coverage of available tumor proteomic datasets necessitated the incorporation of transcriptomic information to expand disease-specific molecular networks. Although our integration strategy prioritized protein-derived associations and downweighted transcriptome-only evidence, the resulting network-wide profiles should not be interpreted as experimentally measured proteomic responses. Furthermore, transcriptomic and proteomic datasets were generally obtained from independent cohorts, introducing potential biological and technical heterogeneity. Third, the scarcity of sufficiently comprehensive, matched compound–proteome perturbation datasets prevented the development of a newly trained end-to-end model capable of directly learning quantitative protein-level responses. Instead, the present framework integrates existing AI-based target predictions with network propagation, producing relative directional influence scores rather than calibrated changes in protein abundance or activity. Fourth, in the absence of systematically validated direct targets for most natural compounds, network propagation was initialized using computationally inferred compound–protein interactions and activation- or inhibition-associated evidence. Consequently, uncertainties in predicted target identities and regulatory directions may propagate through the molecular networks. The herb-level extension is additionally constrained by incomplete information on constituent abundance, bioavailability, pharmacokinetics, and interactions among co-occurring compounds. Importantly, although network propagation substantially expanded molecular coverage, this expansion did not consistently translate into statistically significant disease-specific reversal across cancers after multiple-testing correction. These considerations underscore the need to distinguish computationally inferred network influence from experimentally established molecular perturbation and therapeutic activity.

Collectively, our study establishes a scalable computational foundation for modeling the potential proteome-scale molecular influences of natural products and multicomponent medicinal herbs within disease-specific biological contexts. By connecting standardized chemical knowledge, protein-centered molecular networks, and directional perturbation propagation, this framework moves beyond isolated compound–target associations toward a systems-level representation of natural-product interventions. Future integration of experimentally resolved molecular interactions, quantitative proteomic and phosphoproteomic responses, and concentration- and time-dependent perturbation data could further enable the development of trainable and experimentally calibrated models of complex therapeutic interventions. More broadly, this framework provides a methodological basis for identifying disease-associated intervention modules, prioritizing candidate therapeutic mechanisms, and guiding experimental investigations, offering a path toward more systematic and mechanism-informed natural-product drug discovery.

## Methods

### Collection and curation of natural-product databases

Ingredient and herbal medicine information was collected from three publicly available traditional medicine databases: HERB 2.0, TCMKI, and TCMKD. The original ingredient datasets comprised 44,595, 81,010, and 36,403 records, respectively, yielding 162,008 source records in total. Available annotations included compound names and aliases, molecular formulas, CAS registry numbers, PubChem compound identifiers (CIDs), InChI, InChIKeys, canonical and isomeric SMILES, and database-specific identifiers. Herb identities and herb–ingredient associations were independently extracted from the corresponding source resources.

All source records were retained with their original database identifiers and provenance annotations. Because record completeness and structural consistency varied among databases, chemical identity resolution was performed independently for each source before cross-database integration. Records with insufficient or conflicting evidence were preserved in separate audit categories rather than automatically discarded or assigned potentially incorrect chemical identities. This separation of source-level annotations from standardized chemical entities ensured that subsequent integration could be traced to the original database records.

### Hierarchical PubChem resolution and chemical identity verification

Chemical identities were resolved using a hierarchical, structure-first procedure supported by PubChem PUG REST queries and RDKit-based molecular structure processing. Existing source PubChem CIDs were treated as provisional annotations and verified against available molecular representations, including full InChIKeys, InChI-derived InChIKeys, and isomeric SMILES-derived InChIKeys. Canonical or connectivity-only SMILES were used as supporting evidence but were not considered sufficient to override conflicting stereochemical information.

Resolution was performed in two stages. In the first stage, records were assigned to a strict structural tier when their chemical identities could be established through consistent structure-based evidence, including manually adjudicated cases with resolved source-field conflicts. In the second stage, unresolved records were subjected to conservative structure- and metadata-supported rescue using exact compound names, aliases, CAS numbers, molecular formulas, and available partial structural evidence. Only unambiguous assignments meeting HIGH or MEDIUM HIGH confidence criteria were accepted. Approximate string matching, fuzzy-name reconciliation, and unsupported stereochemical merging were not permitted.

Because the three source databases differed in annotation quality, source-specific adjudication procedures were applied. HERB 2.0 yielded 30,964 structure-resolved and 1,330 additional high-confidence rescued records. TCMKI yielded 30,281 structure-resolved and 11,399 additional accepted records. TCMKD, whose original records already contained PubChem CIDs, underwent extensive structural consistency assessment and conflict correction, yielding 34,670 strict and 1,669 additionally resolved records. The resulting accepted datasets contained 32,294 HERB, 41,680 TCMKI, and 36,339 TCMKD source records.

Records classified as candidate-only, ambiguous, conflicting, or unresolved were retained for provenance and quality-control purposes but excluded from the frozen PubChem-resolved compound registry. PubChem retrieval results were cached locally, and consistency checks were applied to ensure that accepted identifiers agreed with the corresponding chemical annotations.

### Cross-database chemical identity reconciliation and compound registry construction

The accepted PubChem-resolved records from all three databases were combined to construct a unified compound registry. Cross-database identity reconciliation was performed independently of original compound names, using accepted PubChem CIDs and molecular structure information. The 110,313 resolved source records represented 46,423 unique PubChem CIDs and 46,357 distinct full InChIKeys.

To prevent erroneous merging of chemically distinct representations, cases in which multiple PubChem CIDs shared the same full InChIKey were subjected to additional molecular graph analysis. For each candidate pair, PubChem structures were parsed and normalized using RDKit (version 2026.03.5), followed by comparison of stereochemistry-aware molecular graphs, non-stereochemical connectivity, and supporting molecular properties. Among 66 multiple-CID groups sharing a full InChIKey, one pair was classified as exactly graph-equivalent, five as equivalent only after removal of stereochemical information, and 60 as having different molecular graph representations. Only the exactly graph-equivalent pair was merged.

The resulting compound registry contained 46,422 distinct chemical identities. Each entity was assigned a deterministic internal identifier (CMP000001–CMP046422), ordered by the minimum PubChem CID within its accepted identity component. PubChem CIDs were retained as external cross-references rather than permanent internal primary keys. Original database identifiers, source names, resolution methods, confidence tiers, preferred PubChem CIDs, and chemical structure annotations were preserved. Final quality control verified complete source-to-compound mapping, the uniqueness of internal identifiers, consistency of accepted molecular structures, and the absence of prohibited graph-identity merges. Input file hashes and output manifests were retained to support reproducibility.

**Harmonization of herbal medicine identities**

Herbal medicine identities were standardized separately from chemical entities to avoid conflating botanical or medicinal material names with compound-level molecular identities. Source herb information was collected from HERB 2.0, TCMKI, and TCMKD, including primary Chinese names, pinyin, English and Latin names, botanical annotations, medicinal parts, processing information, and database-specific identifiers where available.

Before cross-database harmonization, we collected 23,876 herb entity and name records, including 6,864 entities from HERB, 10,506 from TCMKI, 3,817 from the original TCMKD herb table, and 2,689 additional TCMKD herb-name records that could not be uniquely mapped to its original herb entities. These counts represent source-level records rather than distinct standardized herbs, as equivalent herbs and alternative names may occur across databases. Cross-database equivalence relationships were established conservatively using unique exact matches of normalized primary Chinese names, exact agreement across pinyin, English, and Latin name fields, or manually reviewed equivalence assignments. For previously unresolved TCMKD herb names, candidate mappings were evaluated against HERB and TCMKI annotations and subjected to conservative manual adjudication. Of the 2,689 names examined, 1,841 received accepted identity assignments, whereas 848 remained unresolved and were retained as separate name-based entities. Herb names referring to potentially distinct medicinal parts, processed materials, or preparation forms were not automatically merged solely on the basis of lexical or botanical similarity.

Accepted herb equivalence relationships were represented as a graph, and connected components were used to define unified identities. Components containing multiple distinct source herb entities from the same database were treated as identity conflicts and were not accepted. The final frozen herb registry comprised 12,459 identities with stable identifiers (HRB000001–HRB012459). All original names, identifiers, source membership information, and equivalence evidence were retained for provenance tracking.

**Reconstruction of the unified herb–compound association network**

Original herb–ingredient relationships were reconstructed independently for each source database. HERB 2.0 and TCMKI provided explicit herb-to-ingredient association tables containing 108,337 and 101,001 unique relationships, respectively. For TCMKD, herb associations were extracted from the related herb annotations of ingredient records, producing 88,657 ingredient–herb-name relationships. Together, these resources contained 297,995 source-level herb–ingredient associations.

Each source ingredient was mapped to its corresponding standardized CMP identifier through the frozen compound registry. A total of 274,005 source relationships (91.95%) were successfully linked to accepted chemical identities. Relationships mapping multiple source ingredient records to the same standardized compound within a source herb were collapsed, producing 270,647 source-local herb–compound associations.

Subsequently, source herb identifiers were mapped to the unified HRB registry, and redundant relationships connecting the same HRB and CMP entities were consolidated. This procedure produced 156,588 unique HRB–CMP associations. For each relationship, source database membership, original herb and ingredient identifiers, supporting source records, and the number of supporting databases were preserved. A higher-support subset was defined as associations represented in at least two databases, yielding 81,810 relationships.

All source-to-standardized mappings were subjected to consistency checks, including identification of duplicate associations, unmapped chemical entities, invalid identifiers, and prohibited herb identity merges. The final network was represented as a bipartite herb–compound graph in which HRB and CMP nodes denoted standardized entities and edges denoted curated compositional associations. This representation was used as the chemical and compositional foundation for subsequent compound-level target prediction and herb-level network influence modeling.

**Physicochemical, scaffold and chemical-space characterization**

Physicochemical and structural diversity was characterized using standardized compounds in the final chemical registry. The evaluated molecular descriptors included molecular weight, calculated lipophilicity (cLogP), topological polar surface area, hydrogen-bond donor and acceptor counts, and the fraction of sp3-hybridized carbon atoms (Fsp3). Descriptor distributions were summarized across the compound collection to characterize the physicochemical diversity of the integrated atlas.

The composition of each herb was represented by its set of unique standardized compound identifiers. Herb degree was defined as the number of distinct compounds associated with an herb, whereas compound degree represented the number of distinct herbs containing a given compound. Complementary cumulative degree distributions were used to characterize the heterogeneity of herb composition and compound sharing.

Molecular scaffold assignments were used to quantify structural richness within individual herbs. Scaffold richness was defined as the number of distinct molecular scaffolds represented among the compounds associated with an herb. Spearman rank correlation was used to assess the relationship between compound count and scaffold richness. To reduce the effect of unequal constituent numbers, scaffold diversity was additionally evaluated through rarefaction to a common sample size of 20 cyclic compounds per herb.

Compound-level chemical space was visualized using uniform manifold approximation and projection (UMAP), with points colored according to cLogP. For comparisons between herbs, compound overlap and scaffold overlap were quantified using Jaccard similarity. Compound Jaccard similarity was defined as the intersection divided by the union of standardized compound sets, whereas scaffold Jaccard similarity was defined analogously using molecular scaffold sets. Pairwise structural similarity was further evaluated using molecular fingerprint-based Tanimoto similarity, with herb pairs categorized according to whether they shared exact compounds, shared scaffolds without exact compound overlap, or shared neither.

**Collection and harmonization of cancer multi-omics datasets**

Transcriptomic and proteomic datasets were assembled for nine cancer types, including STAD, BRCA, KIRC, LUAD, LUSC, LIHC, HNSC, COAD, and UCEC. RNA-sequencing data were obtained from TCGA through the Genomic Data Commons (GDC), together with normal tissue reference data from GTEx. Quantitative proteomic profiles and associated clinical metadata were retrieved from cancer-specific studies deposited in the Proteomic Data Commons (PDC). Sample annotations were harmonized to distinguish primary tumors, adjacent non-malignant tissues, and independent normal tissue references.

Three distinct transcriptomic data branches were maintained according to their analytical purposes. First, harmonized TCGA–GTEx gene expression data processed through the Toil pipeline and distributed through UCSC Xena were used for differential expression analyses. Second, TCGA STAR unstranded read counts obtained through GDC were used to reconstruct tumor- and adjacent-normal transcriptomic association networks. Third, GTEx v8 gene-level RNA-sequencing data were processed independently to construct tissue-matched normal reference networks. These datasets were not pooled indiscriminately, because their processing pipelines, study origins, and sample compositions differed.

For each cancer type, sample identifiers, tissue classifications, case identifiers, sequencing files, and repeated sample representations were audited before analysis. Where repeated samples or files occurred within the same case and tissue class, deterministic metadata- and quality-control-based selection procedures were applied. Proteomic samples were independently annotated according to their tumor or non-malignant status and case identifiers. Cohort-specific information, including tumor–normal pairing, repeated observations, tissue proxies, and differences between transcriptomic and proteomic populations, was preserved throughout processing.

Importantly, transcriptomic and proteomic measurements generally originated from independent cohorts. Accordingly, multi-omics integration was performed at the molecular network level rather than through sample-matched transcript–protein measurements. Cross-study normal reference comparisons were interpreted as cross-sectional molecular state comparisons and not as direct longitudinal observations of disease progression.

**Transcriptomic preprocessing and differential gene expression analysis**

Differential gene expression was analyzed using harmonized TCGA and GTEx transcriptomic data obtained from the UCSC Xena Toil resource. The source expression values represented log2-transformed RSEM expected counts. Count-scale values were reconstructed using round ($2^x - 1$), where x denotes the stored log2-transformed expression value.

For each cancer, tumor samples and appropriate non-malignant reference samples were selected according to the predefined tissue crosswalk. Differential expression analysis was performed using DESeq2[34]. A negative binomial generalized linear model incorporating study origin and disease status was fitted using the design formula ~ study + status, allowing the principal tumor–normal contrast to be estimated while accounting for study-associated differences between TCGA and GTEx. Log2 fold-change estimates were stabilized using apeglm shrinkage[35]. Multiple-testing correction was performed using the Benjamini–Hochberg procedure.

Differential expression results were retained at the gene level, including estimated log2 fold changes, associated statistical significance, and directional classifications. Gene identifiers were harmonized to canonical human gene symbols, with ambiguous or nonunique mappings excluded where necessary. These signatures were used to characterize cancer-specific transcriptional alterations and to define disease-associated molecular reference sets for downstream analyses.

For network reconstruction, GDC STAR unstranded raw counts were processed separately from the Xena differential expression branch. Within each cancer, tumor and adjacent-normal samples were jointly subjected to DESeq2 variance-stabilizing transformation using a condition-based design with blind = FALSE. The transformed expression matrix was subsequently separated by tissue state. State-specific correlation networks were constructed from variance-stabilized expression values rather than raw read counts.

GTEx v8 reference data were processed as an independent normal-tissue expression branch and aligned to the same gene identifier and PPI reference space. These reference samples were not mixed into the original TCGA tumor–adjacent correlation estimation. Their distinct cohort origin was retained when interpreting normal-reference networks.

**Proteomic data processing and differential protein abundance analysis**

Quantitative tumor and non-malignant tissue proteomic datasets were obtained from cancer-specific PDC studies. Protein abundance matrices, sample annotations, and available case-level metadata were harmonized within each cohort. Protein identifiers were mapped to canonical human gene symbols, and duplicated or ambiguous protein-to-gene assignments were assessed before downstream analysis.

The network reconstruction branch used normalized log2-ratio protein abundance values supplied by the processed proteomic datasets. As these measurements were continuous quantitative abundance profiles rather than sequencing counts, count-based normalization methods such as DESeq2 variance-stabilizing transformation were not applied. Protein-level filtering was performed separately for differential abundance analysis and network correlation estimation, according to their respective requirements.

Differential protein abundance was analyzed using the limma framework[36]. Tumor and non-malignant samples were compared using linear models that retained all valid observations. Where repeated measurements or tumor–normal pairs were available, case-level dependence was incorporated through a blocking strategy and duplicateCorrelation when supported by the study design. This approach allowed paired cases and valid singleton samples to contribute to the analysis without restricting the entire cohort to complete pairs. Empirical Bayes moderation was used for differential abundance estimation, with the availability of robust estimation and the fitted correlation structure assessed for each cohort.

The resulting protein-level statistics included estimated log2 abundance differences, significance measures, and directional classifications of differential protein abundance. These results were used to characterize proteomic alterations across cancer types and to complement transcriptomic disease signatures. Importantly, differential protein abundance estimates were not substituted for transcriptomic effects or assumed to correspond directly to changes in protein activity.

**Integration of physical protein–protein interaction networks**

A reference human protein–protein interaction network was constructed by integrating physical interaction information from STRING, BioGRID, and IntAct. Interacting proteins were mapped to canonical human gene identifiers, and interactions were represented as unordered protein pairs. Self-interactions, redundant records, and incompatible identifier mappings were excluded during reference network construction.

The integrated reference was represented as an undirected and unsigned physical interaction scaffold, containing 18,908 proteins and 1,053,426 unique protein–protein interactions. This broad canonical network provided a common set of biologically supported candidate interactions for subsequent transcriptomic and proteomic network reconstruction.

For each cancer and molecular modality, the reference interactome was restricted to proteins with eligible expression measurements. Transcriptomic and proteomic networks were initially reconstructed independently, using modality-specific node coverage and corresponding subsets of the common PPI scaffold. Within a given modality, tumor and non-malignant states were constrained to a shared candidate node and interaction universe, ensuring that subsequent comparisons were not driven solely by differences in gene or protein detectability.

Importantly, the physical PPI scaffold itself did not encode regulatory activation or inhibition. Signed interaction weights were inferred subsequently from state-dependent molecular associations. The sign of an association therefore represented positive or negative statistical coupling, rather than direct evidence of a biochemical activating or inhibitory mechanism.

**Construction of state-specific PPI-restricted molecular association networks**

For each cancer type, transcriptomic and proteomic molecular association networks were estimated independently within the common physical protein–protein interaction (PPI) scaffold. Gene identifiers from RNA-sequencing data and protein identifiers from quantitative proteomics were mapped to canonical human gene symbols. Ambiguous or duplicated mappings were excluded before determining the eligible node set. Within each modality, tumor and non-malignant states were constrained to the same retained PPI-supported node and candidate-edge universe, so that differences in network weights were not confounded with changes in assay coverage.

RNA associations were computed from DESeq2 variance-stabilized expression profiles, using tumor and adjacent-normal samples processed jointly before separation by state. For proteomics, the input consisted of normalized log2-ratio abundance profiles rather than sequencing counts. Proteins were required to have non-missing values in at least 80% of samples in each state, variance greater than $10^{-12}$ within both states, an unambiguous gene-symbol assignment and at least one retained PPI interaction. Pairwise-complete Pearson correlations were calculated without global abundance imputation; protein–protein edges were retained only if at least 80% pairwise sample coverage was achieved in each compared state.

For a PPI-supported pair of molecular features i and j in cancer c, modality m and state s, the observed association was defined as:

$$r_{ij}^{(c,m,s)} = \mathrm{corr}\left(x_i^{(c,m,s)}, x_j^{(c,m,s)}\right)$$

Only candidate edges present in the PPI scaffold were assigned nonzero correlation weights. The symmetric observed adjacency was therefore defined by the PPI indicator function:

$$A_{ij}^{(c,m,s)} = \mathbb{1}_{\left[(i,j) \in E_{PPI}^{(c,m)}\right]} r_{ij}^{(c,m,s)}$$

Self-loops were set to zero. Positive and negative correlations were interpreted as concordant and opposing statistical associations of molecular abundances, respectively not as direct biochemical activation or inhibition. Observed signed edge tables and matrices were preserved for deconvolution and quality control.

**Spectrally constrained network deconvolution**

Because observed molecular correlations may reflect both direct associations and dependencies mediated by intermediate proteins, network deconvolution (ND) was applied separately to the RNA- and protein-derived association matrices[37]. For each cancer, modality and state, the PPI-masked observed adjacency A was scaled by a positive coefficient α and transformed as follows:

$$B = \alpha A; \quad D = B(I + B)^{-1} = I - (I + B)^{-1}$$

Equivalently, each eigenvalue μ of the observed matrix A is mapped to the corresponding deconvolved eigenvalue:

$$\mathrm{d}(\mu) = \frac{\alpha\mu}{1 + \alpha\mu}$$

To avoid unstable transformation near singular spectral regions, an admissible scaling factor was derived from the largest positive and the most negative eigenvalues of each observed matrix. Using a spectral control parameter β = 0.5, the state-specific upper bound was:

$$\alpha_{max,s} = \min\left(\frac{\beta}{(1 - \beta)\mu_{max,s}}, \frac{\beta}{(1 + \beta)|\mu_{min,s}|}\right)$$

A single α was then shared across the contrasted states within the same cancer and modality, calculated as the minimum of the state-specific admissible bounds:

$$\alpha_{shared} = \min_s \alpha_{max,s}$$

The deconvolution matrix was computed from the complete PPI-masked observed association matrix, after which only the weights corresponding to retained physical PPI edges were extracted. Thus, the matrix inversion could account for indirect paths, but the exported adjacency remained confined to the PPI scaffold; non-PPI entries were not constrained to remain zero during the inversion itself. Symmetry, diagonal structure, finite weights and edge-table consistency were verified after transformation. For the reference STAD analysis, the shared scaling factors were 0.005387536155 for RNA and 0.0105987565 for protein. Other cancer-specific factors were determined by the same spectral procedure rather than copied from STAD.

**Bootstrap estimation of transcriptomic association stability**

Because RNA-derived correlations can be sensitive to sampling variation, particularly when the number of non-malignant specimens is limited, the stability of their signs was estimated by nonparametric bootstrap resampling. Within each cancer and tissue state, n biological samples were resampled with replacement B = 1,000 times. For each bootstrap replicate, Pearson correlations were recalculated on the eligible PPI-supported RNA edges. The reference STAD implementation used random seed 20260929. Bootstrap results were summarized by sign counts and did not replace the full-data deconvolved edge magnitudes.

For each edge, q was defined as the proportion of bootstrap correlations retaining the sign of the full-data observed RNA correlation:

$$q_{ij} = \frac{1}{B}\sum_{b=1}^{B} \mathbb{1}\left[\mathrm{sgn}\left(r_{ij}^{(b)}\right) = \mathrm{sgn}\left(r_{ij}^{(obs)}\right)\right]$$

Bootstrap support was mapped to an RNA confidence weight in the range from zero to one, where chance-level sign support contributed no confidence:

$$C_{ij}^{R} = \max\left(0, 2q_{ij} - 1\right)$$

Because the integrated network used ND-derived associations whereas bootstrap support was estimated for raw correlations, an additional concordance check compared raw and deconvolved RNA signs. The effective confidence was set to zero whenever these signs disagreed:

$$C_{ij}^{R,eff} = C_{ij}^{R} \cdot \mathbb{1}\left[\mathrm{sgn}\left(r_{ij}^{(obs)}\right) = \mathrm{sgn}\left(D_{ij}^{R}\right)\right]$$

Effective RNA confidence was used only to modulate the RNA contribution during cross-omics integration. Protein deconvolution scores served as network anchors and were not subjected to a separate bootstrap-derived reweighting step in the frozen integration procedure.

**Cross-omics calibration and protein-centered signed network integration**

RNA and proteomic deconvolved weights were integrated using a protein-centered, RNA-corrected design. To account for differences in the original weight scales, the absolute ND weights of tumor and non-malignant networks were pooled within each modality and converted to empirical rank distributions; the original signs were then restored. Pooling the compared states within a modality retained their relative difference in weight magnitude, rather than normalizing each state independently. The resulting normalized scores were:

$$R_{ij} = \mathrm{sgn}\left(D_{ij}^{R}\right)F_{R}\left(\left|D_{ij}^{R}\right|\right); \quad P_{ij} = \mathrm{sgn}\left(D_{ij}^{P}\right)F_{P}\left(\left|D_{ij}^{P}\right|\right)$$

Here, $F_R$ and $F_P$ denote the pooled within-modality empirical rank transformations; the state and cancer indices are suppressed for clarity. RNA–protein agreement was quantified across their common edges using weights determined by the effective RNA confidence and the geometric mean of the respective absolute normalized scores:

$$\omega_{ij} = C_{ij}^{R,eff}\sqrt{\left|P_{ij}\right|\left|R_{ij}\right|}$$

Weighted observed sign agreement and its expectation from marginal sign frequencies were computed as follows. In the expected-agreement expression, the second subscript p or n denotes a positive or negative sign, respectively, and the first subscript denotes the protein (P) or RNA (R) modality:

$$p_o = \frac{\sum_{ij}\omega_{ij}\mathbb{1}\left[\mathrm{sgn}\left(P_{ij}\right) = \mathrm{sgn}\left(R_{ij}\right)\right]}{\sum_{ij}\omega_{ij}}$$

$$p_e = p_{\mathrm{P,p}}p_{\mathrm{R,p}} + p_{\mathrm{P,n}}p_{\mathrm{R,n}}$$

A nonnegative chance-corrected agreement coefficient was then used to derive the integration parameters:

$$K_{RP} = \max\left(0, \frac{p_o - p_e}{1 - p_e}\right); \quad \lambda = \frac{K_{RP}}{2}; \quad \kappa = K_{RP}$$

To prevent differences in fusion strength from creating artificial state-specific contrasts, parameter estimation pooled the compared tissue states for each cancer. The network was divided into three edge classes. On edges observed by both modalities, the protein score P was retained as the anchor and corrected according to RNA sign agreement and stability:

$$W_{ij,both} = P_{ij}\left[1 + \lambda\,\mathrm{sgn}\left(P_{ij}R_{ij}\right)\left|R_{ij}\right|C_{ij}^{R,eff}\right]$$

Concordant RNA evidence increased protein-edge magnitude, whereas discordant RNA evidence attenuated it; the calibrated correction was constrained not to reverse the protein-derived sign. Protein-only and RNA-only edges were assigned as follows:

$$W_{ij,P-only} = P_{ij}; \quad W_{ij,R-only} = \kappa R_{ij}C_{ij}^{R,eff}$$

Accordingly, RNA-only edges expanded coverage beyond the measured proteome but were explicitly attenuated, whereas protein evidence dominated shared and protein-only interactions. Parameters were derived exclusively from network-level RNA–protein concordance and were not optimized against drug-response results, disease reversal, survival outcomes or independent perturbation validations. These numerical values illustrate the reference implementation, and each cancer was calibrated under the same rules using its own molecular data.

**Final network assembly, structural quality control and signed transition construction**

Integrated adjacency matrices were assembled separately for tumor and non-malignant conditions from the union of retained RNA and protein interaction edges. Within each cancer, a shared molecular node universe and stable node ordering were enforced across the compared states. Zero-weight edges were distinguished from active interactions, and connected components and isolated nodes were audited before freezing the network. Where necessary, a common connected node set was selected to ensure that downstream state comparisons used the same propagation space. This procedure retained a disease-state-dependent weighted and signed network while maintaining common molecular identifiers and a consistent physical PPI prior.

For the STAD reference network, the initial union comprised 16,901 nodes and 998,343 candidate edges. Two RNA-only nodes (KLHL30 and PLAC4) had no effective connections in the tumor state and were removed from both states by using the tumor largest connected component as the common node set. The final shared network contained 16,899 nodes and 998,341 retained candidate edges, of which 995,799 were active in tumor and 994,978 in adjacent-normal tissue. Final quality control checked matrix symmetry, zero diagonal, matching node orders, connectedness, zero-degree nodes, degree and absolute-strength distributions, as well as agreement between edge tables and sparse adjacency matrices. Signed weights characterize state-dependent statistical coupling and should not be interpreted directly as causal activation or inhibition.

For downstream SRWR, each finalized signed adjacency W was decomposed into its nonnegative positive and negative components:

$$W_{ij}^{(+)} = \max(W_{ij}, 0); \quad W_{ij}^{(-)} = \max(-W_{ij}, 0)$$

The absolute weighted degree of node i was defined by the total magnitude of its incident signed interactions:

$$d_i = \sum_j |W_{ij}|$$

Positive and negative transition probabilities were normalized using the same absolute degree, preserving total probability mass over both interaction signs:

$$P_{ij}^{(+)} = \frac{W_{ij}^{(+)}}{d_i}; \quad P_{ij}^{(-)} = \frac{W_{ij}^{(-)}}{d_i}$$

$$\sum_j \left(P_{ij}^{(+)} + P_{ij}^{(-)}\right) = 1$$

The final positive and negative transition matrices were retained as fixed derivatives of the reconstructed molecular network. Their row sums, signed reconstruction, and numerical consistency were verified before the subsequent propagation analysis; no network weights were modified during this transition-matrix construction.

**Compound and human protein representations for proteome-wide target inference**

Proteome-wide compound–protein interaction inference was performed using the standardized compound registry and a unified human protein target panel. The compound registry contained 46,422 chemically resolved entities, of which 46,421 were included in the completed large-scale inference matrices. The target panel comprised 19,281 human protein-coding genes represented by canonical HGNC gene symbols and their corresponding protein sequences. Protein identifiers were harmonized across prediction models, while the original protein sequence and accession mappings were retained for traceability.

Molecular representations were generated according to the input requirements of each predictive framework. For ConPLex[38], compounds were represented using molecular fingerprints, and proteins were encoded using ProtBert-derived sequence embeddings. For DTIAM[39], molecular representations were obtained using the pretrained BERMol framework, whereas protein representations were generated using ESM-2 sequence embeddings. Precomputed compound and protein representations were reused throughout large-scale inference to avoid redundant feature extraction. Proteins sharing identical amino acid sequences were retained as distinct gene-level targets in the final prediction space, while sequence-level feature redundancy was handled through reusable representation caches.

The resulting common compound and protein identifier spaces enabled model-specific interaction predictions to be compared across the same human target panel. All compound–protein prediction outputs were linked to stable internal compound identifiers and canonical target gene symbols, allowing subsequent integration with the disease-specific molecular networks.

**Large-scale drug–target interaction prediction and weighted cross-model consensus**

Compound–protein interaction inference was carried out using the frozen compound registry and a harmonized panel of 19,281 human protein targets. Three completed, complementary prediction configurations were used: ConPLex (BindingDB), DTIAM-Merged and DTIAM-Yamanishi. The incomplete DTIAM-Hetionet inference branch was excluded from the production workflow, all consensus calculations and subsequent analyses. Compound and target identifiers were aligned across model outputs before combining their predictions. The inference panel was evaluated independently of cancer-specific expression data so that the resulting compound–protein evidence could subsequently be transferred, without refitting, to different disease networks.

For each compound, model-specific candidate targets were selected by ranking prediction scores within the common human protein panel and retaining the top 1% (193 targets). For DTIAM-Merged and DTIAM-Yamanishi, a raw interaction-score threshold of at least 0.5 was additionally required. ConPLex scores were interpreted as relative similarity scores for ranking rather than as calibrated binding probabilities. The binary support of prediction model m for compound c and protein target g was defined as:

$$b_{\mathrm{cgm}} = \mathbb{1}[\text{model m retains target g for compound c}]$$

The support indicators from the three configurations were combined using fixed model weights of 0.4, 0.4 and 0.2, respectively, yielding the weighted consensus score:

$$a_{\mathrm{cg}} = 0.4 b_{\mathrm{cg,ConPLex}} + 0.4 b_{\mathrm{cg,Merged}} + 0.2 b_{\mathrm{cg,Yamanishi}}$$

Accordingly, the possible nonzero support scores were 0.2, 0.4, 0.6, 0.8 and 1.0. Candidate pairs were summarized at cumulative support thresholds of 0.4, 0.6, 0.8 and 1.0, whereas the distinct exact-support groups were used to examine regulatory-mode evidence. The support score denotes the summed weight of models retaining a candidate interaction, not a calibrated probability that the compound binds the protein. This frozen consensus layer was retained separately from subsequent activation–inhibition inference and disease-specific network propagation.

**Activation–inhibition inference and characterization of directional target landscapes**

Potential regulatory modes of the predicted compound–protein pairs were evaluated using separate DTIAM-based activation and inhibition models. For each pair, activation-associated and inhibition-associated scores, denoted by $A_{cg}$ and $I_{cg}$, were recorded independently. These values represent model-derived evidence for potential modes of action and were not interpreted as measured biochemical activation or inhibition, target occupancy, or protein-expression fold changes. In particular, strong interaction-model support does not itself determine the direction of a biological effect.

To describe the global regulatory landscape, high activation and inhibition evidence was defined using the 95th-percentile cutoff of each corresponding model's complete score distribution. The descriptive high-score indicators were:

$$H_{\mathrm{cg}}^{\mathrm{A}} = \mathbb{1}\left[A_{\mathrm{cg}} \geq \mathrm{Q}_{0.95}(A)\right]$$

$$H_{\mathrm{cg}}^{\mathrm{I}} = \mathbb{1}\left[I_{\mathrm{cg}} \geq \mathrm{Q}_{0.95}(I)\right]$$

Each pair was assigned to one of four mutually exclusive evidence categories: activation-high only, inhibition-high only, both high or neither high. Category proportions were compared between exact weighted-DTI support groups. Joint activation–inhibition score distributions were characterized both across the complete prediction universe and among pairs meeting the weighted support threshold of 0.4. The both-high category was preserved as evidence of potentially ambiguous or context-dependent regulatory mode, rather than forcibly reduced to a single activation or inhibition label.

At the compound level, direction-resolved target counts and the number of compounds achieving progressively larger minimum target counts were summarized through empirical cumulative distributions and coverage curves. Directional composition was assessed from the relative number of activation-favored and inhibition-favored targets among covered compounds. These analyses characterized heterogeneity in the predicted regulatory landscape; critically, the descriptive global 95th-percentile thresholds were not used as the signed-seed eligibility rule for SRWR. All downstream propagation used the independently frozen seed table defined below.

**Construction of frozen signed compound–protein perturbation seeds**

A fixed signed compound–protein interaction resource was constructed from the weighted DTI consensus and the two regulatory-mode scores. A pair was eligible for propagation when the weighted DTI support was at least 0.4 and at least one of two directional criteria was satisfied: (i) the absolute activation–inhibition score difference was at least 0.20; or (ii) a canonical one-high/one-low pattern was present, with one regulatory score greater than 0.5 and the opposing score less than 0.5. Thus, the eligibility indicator was:

$$E_{\mathrm{cg}} = \mathbb{1}\left[a_{\mathrm{cg}} \geq 0.4\right] \cdot \mathbb{1}\left[\Delta_{\mathrm{cg}} \geq 0.20 \ \vee \ C_{\mathrm{cg}} = 1\right]$$

Here, $\Delta_{\mathrm{cg}} = |A_{\mathrm{cg}} - I_{\mathrm{cg}}|$ denotes directional-score separation, and the canonical one-high/one-low indicator is:

$$C_{\mathrm{cg}} = \mathbb{1}\left[\left(A_{\mathrm{cg}} > 0.5 \ \wedge \ I_{\mathrm{cg}} < 0.5\right) \ \vee \ \left(I_{\mathrm{cg}} > 0.5 \ \wedge \ A_{\mathrm{cg}} < 0.5\right)\right]$$

For eligible pairs, a signed seed weight was formed by multiplying the weighted interaction support by the difference between activation and inhibition scores:

$$z_{\mathrm{cg}} = a_{\mathrm{cg}}\left(A_{\mathrm{cg}} - I_{\mathrm{cg}}\right); \quad E_{\mathrm{cg}} = 1$$

Positive $z_{cg}$ values represented activation-like initial perturbations, and negative values represented inhibition-like initial perturbations. Seed magnitudes jointly reflected cross-model DTI support and separation of the two regulatory scores; they were not treated as measured effects on protein activity. Pairs that failed either eligibility component were not used as direct network-perturbation seeds. The production resource preserved the original model scores, selection reasons, regulatory-direction labels and final signed seed values. For all STAD and pan-cancer propagation analyses, the frozen values were used directly and were not recalculated, threshold-tuned or reweighted according to disease signatures, network results or compound-validation outcomes.

**Construction of disease-specific signed transition matrices**

Signed propagation was performed on the protein-centered, RNA-corrected cancer molecular networks described in the preceding sections. Each tumor-state network was represented by a symmetric weighted adjacency matrix $W$, whose positive and negative values encoded concordant and opposing statistical molecular associations, respectively. The propagation calculations did not equate these signs with experimentally verified activating and inhibitory PPI mechanisms. The final connected STAD tumor network contained 16,899 protein-associated nodes and served as the primary example.

For each signed network, the adjacency was separated into nonnegative positive and negative components:

$$W_{ij}^{(+)} = \max(W_{ij}, 0); \quad W_{ij}^{(-)} = \max(-W_{ij}, 0)$$

The absolute weighted degree of node *i* was defined by the sum of the magnitudes of its incident edges:

$$d_i = \sum_j |W_{ij}|$$

The positive and negative transition probabilities were normalized by this same absolute degree:

$$T_{ij}^{(+)} = \frac{W_{ij}^{(+)}}{d_i}; \quad T_{ij}^{(-)} = \frac{W_{ij}^{(-)}}{d_i}$$

For every connected node, the outgoing probability mass satisfies:

$$\sum_j \left(T_{ij}^{(+)} + T_{ij}^{(-)}\right) = 1$$

To preserve perturbation sign during propagation, two channels were assigned to each protein: one for positive and one for negative influence. Positive edges preserved the signal channel, whereas negative edges switched it. The corresponding lifted transition matrix was:

$$T_{lift} = \begin{bmatrix} T^{(+)} & T^{(-)} \\ T^{(-)} & T^{(+)} \end{bmatrix}$$

This matrix is row stochastic and has dimension $2N \times 2N$ for a network containing $N$ connected molecular nodes (33,798 × 33,798 for STAD). The two channels represent computational perturbation states, not twice as many distinct proteins. The transition matrices were generated as fixed derivatives of the reconstructed network without modifying its original edge weights, and their row sums and signed-weight reconstruction were checked before SRWR inference.

**Signed random walks with restart and proteome-wide influence-profile generation**

For each natural compound, the frozen signed compound–protein seeds were mapped to the HGNC gene symbols of the tumor-state network. Compounds were considered ineligible when they lacked any retained signed DTI seed or when no signed target could be mapped to the corresponding cancer network. Eligibility was reassessed in each cancer, because networks differed in node coverage, while the compound–protein seed weights themselves were held fixed.

For an eligible compound *c*, the mapped signed seeds $z_{cg}$ were separated into positive and negative restart channels:

$$s_{cg}^{(+)} = \max(z_{cg}, 0); \quad s_{cg}^{(-)} = \max(-z_{cg}, 0)$$

The channel-stacked restart vector was normalized by the total absolute mass of the mapped signed seeds:

$$S_c = \frac{1}{\sum_{g \in V_c} |z_{cg}|} \begin{pmatrix} s_c^{(+)} \\ s_c^{(-)} \end{pmatrix}; \quad \|S_c\|_1 = 1$$

where $V_c$ denotes the molecular node universe of the selected cancer network. The iterative signed random walk was initialized at the restart vector, $R_c^{(0)} = S_c$, and updated as:

$$R_c^{(t+1)} = (1 - \alpha) T_{lift}{}^{T} R_c^{(t)} + \alpha S_c$$

The restart probability was fixed at $\alpha = 0.3$. Iteration stopped when the L1 difference between successive state vectors was less than $10^{-10}$ or when 500 iterations had been reached:

$$\left\| R_{c}^{(t+1)} - R_{c}^{(t)} \right\|_{1} < 10^{-10}$$

Calculations were performed in batches of 64 compounds using float64 arithmetic for the propagation iterations. Completed influence profiles were saved as float32 arrays, together with compound identifiers, gene order, mapped seed counts, raw absolute seed mass, status, iteration number, final residual and convergence indicators. To obtain a single signed influence profile, the two channels at convergence were subtracted:

$$Y_{cg} = R_{cg}^{(+)} - R_{cg}^{(-)}$$

A positive $Y_{cg}$ indicates positive network-derived influence and a negative value indicates negative network-derived influence under the signed statistical interaction model. Scores are dimensionless outputs of network propagation; they are neither experimentally observed protein-abundance changes nor quantitative predictions of transcriptional fold change, target occupancy or drug concentration. In the STAD reference run, 45,418 eligible compounds yielded frozen influence profiles across 16,899 network nodes. The same algorithm, restart parameter, convergence criteria and numerical implementation were used for the eight additional cancer networks, with cancer-specific mappings and network adjacency matrices.

**Quantification of direct-to-network scope and disease-opposing influence**

To distinguish network-wide inferential coverage from direct compound–target predictions, we quantified the number of signed direct targets and the number of network nodes assigned nonzero propagated scores. For compound *c*, direct scope was defined as the count of its distinct mapped signed targets, whereas propagated scope was the count of nodes with nonzero SRWR influence:

$$N_{c}^{\text{direct}} = \sum_{g \in V_c} \mathbb{1}\left[z_{cg} \neq 0\right]$$

$$N_{c}^{\text{SRWR}} = \sum_{g \in V_c} \mathbb{1}\left[Y_{cg} \neq 0\right]$$

For compounds with at least one mapped signed direct target, the expansion ratio was:

$$E_{c}^{\text{network}} = \frac{N_{c}^{\text{SRWR}}}{N_{c}^{\text{direct}}}$$

The distribution of this ratio was summarized over all STAD compounds with completed network propagation. Because diffuse random-walk influence can be numerically small yet nonzero, this measure quantifies the scope of assigned network scores rather than the number of experimentally responsive or confidently affected proteins.

We additionally evaluated network-derived influence within three independently defined STAD disease signatures: the transcriptomic differential-expression set, the proteomic differential-abundance set, and the RNA–protein concordant core. A disease-opposing influence was defined as a negative score for a disease-upregulated molecular feature or a positive score for a disease-downregulated feature. For a signature $D$ partitioned into $D_{UP}$ and $D_{DOWN}$, the propagated disease-opposing scope was:

$$N_{c,D}^{\text{opp,SRWR}} = \sum_{g \in D_{UP}} \mathbb{1}\left[Y_{cg} < 0\right] + \sum_{g \in D_{DOWN}} \mathbb{1}\left[Y_{cg} > 0\right]$$

The corresponding direct-target scope used the same definition with signed direct seeds $z_{cg}$ in place of $Y_{cg}$:

$$N_{c,D}^{\text{opp,direct}} = \sum_{g \in D_{UP}} \mathbb{1}\left[z_{cg} < 0\right] + \sum_{g \in D_{DOWN}} \mathbb{1}\left[z_{cg} > 0\right]$$

For compound–signature pairs with nonzero directly assigned disease-opposing scope, the directional expansion ratio was:

$$E_{\mathrm{c,D}}^{\mathrm{opp}} = \frac{N_{\mathrm{c,D}}^{\mathrm{opp,SRWR}}}{N_{\mathrm{c,D}}^{\mathrm{opp,direct}}}$$

The analyses were performed separately for the RNA, protein and concordant-core signatures, with 36,812, 26,418 and 14,114 eligible compound–signature comparisons, respectively. These are numbers of evaluable compound-level comparisons rather than numbers of distinct genes in the three signatures. Scope expansion was evaluated independently of the subsequent matched-compound null tests, pathway analyses and external drug-perturbation experiments, which address different questions about disease specificity and biological validity. No increase in prediction accuracy or therapeutic efficacy was inferred from scope expansion alone.

**Computational reproducibility and quality control**

All compound–protein predictions, directional inference outputs, signed seeds, and network propagation results were retained with stable compound and protein identifiers. Model-specific outputs were audited for compound coverage, target identifier consistency, finite prediction values, completed inference records, and agreement between source prediction matrices and downstream integrated tables.

The weighted DTI consensus and directional perturbation seed tables were frozen before subsequent cancer-specific network propagation and biological validation. Model outputs, consensus weights, regulatory-direction scores, selection criteria, and final signed-seed values were reused unchanged across cancer types. In particular, compound-specific signed seeds were not recalibrated using disease expression signatures or downstream molecular recovery outcomes.

For each cancer-specific propagation run, quality control recorded the number of eligible signed targets, the number successfully mapped to the network, the total absolute seed weight, the fraction of mapped seeds, propagation status, convergence iterations, final L1 residuals, and numerical validity of the resulting signed profiles. Sparse signed transition matrices were separately checked for normalization consistency, finite weights, and agreement with the frozen network adjacency.

Completed prediction profiles were stored in structured HDF5 matrices indexed by compound identifiers and canonical network gene symbols. These retained the complete network-wide signed profiles and selected high-ranking network effects, enabling subsequent analyses to use identical frozen prediction outputs without repeated DTI inference or retrospective adjustment of the propagation procedure.

**Literature-supported compound curation and matched-control construction**

A literature-supported compound collection was curated to examine whether compounds with reported activity against gastric cancer exhibited preferentially disease-opposing network influence in stomach adenocarcinoma (STAD). Candidate compounds were cross-referenced against the chemically standardized registry while preserving their original evidence records and compound identities. The source collection contained 35 literature-supported compounds. One compound (CMP011784) had no eligible signed drug–target interaction (DTI) seed and could not produce an STAD-specific signed random walk with restart (SRWR) profile; the evaluable literature set therefore comprised 34 compounds.

The fixed STAD tumor-state network contained 16,899 molecular nodes, and the previously generated SRWR profiles were used without recalculating target predictions or selecting compounds according to validation outcomes.

To control for systematic differences in the direct-target evidence available for individual compounds, matched controls were sampled from the frozen compound registry using three covariates: the total number of signed DTI targets, the number of signed targets mapped to the STAD network, and the total absolute weight of network-mapped signed DTI seeds. For compound $c$, the matching vector was

$$\mathbf{x}_c = \left(\log\left(1 + n_c^{\text{signed}}\right),\ \log\left(1 + n_c^{\text{mapped}}\right),\ \log\left(1 + M_c^{\text{seed}}\right)\right).$$

The mapped seed mass was defined as

$$M_c^{\text{seed}} = \sum_{g \in V_{\text{STAD}}} |z_{cg}|,$$

where $z_{cg}$ is the frozen signed DTI seed and $V_{\text{STAD}}$ denotes the STAD network node set. All covariates were robustly standardized using their medians and interquartile ranges. Candidate controls were ordered by proximity in the resulting three-dimensional matching space, and a fixed pool of the 50 nearest eligible compounds was used for each literature compound. All 44 members of the broader gastric-cancer validation candidate collection were excluded from the control population to prevent potential case contamination. With a fixed random seed of 20261006, 10,000 matched control sets were constructed, each containing 34 distinct compounds. The resulting literature-plus-control universe comprised 1,682 unique compounds, for which the same frozen 16,899-node SRWR profiles were evaluated in all subsequent analyses. Control assignments were not regenerated between endpoints.

**Disease-associated molecular signatures and the concordant RNA–protein core**

STAD molecular disease signatures were derived from the transcriptomic and proteomic differential analyses described in the preceding Methods sections. Molecular identifiers were harmonized to the common STAD network gene universe. Significance was determined using a Benjamini–Hochberg-adjusted $P < 0.05$, with disease direction assigned from the apeglm-shrunken log2 fold change for RNA and the limma log2 fold change for protein abundance. The resulting RNA signature contained 4,501 tumor-upregulated and 5,761 tumor-downregulated genes (10,262 total), whereas the protein signature contained 2,939 upregulated and 2,844 downregulated proteins (5,783 total).

The cross-omics concordant core was restricted to features significantly altered in both assays and agreeing in the direction of the tumor-associated change. Upregulated and downregulated core sets were defined, respectively, as

$$D_{\text{Core}}^{\text{UP}} = D_{\text{RNA}}^{\text{UP}} \cap D_{\text{Protein}}^{\text{UP}},$$
$$D_{\text{Core}}^{\text{DOWN}} = D_{\text{RNA}}^{\text{DOWN}} \cap D_{\text{Protein}}^{\text{DOWN}}.$$

This yielded 1,478 concordantly upregulated and 1,586 concordantly downregulated genes (3,064 core features). Within each omics modality and disease-direction group, features were ranked independently by the absolute magnitude of the differential effect, with rank 1 indicating the strongest effect. For the core, a joint disease-strength ranking was obtained from the arithmetic mean of the corresponding within-direction RNA and protein ranks:

$$r_g^{\text{Core}} = \frac{r_g^{\text{RNA}} + r_g^{\text{Protein}}}{2}.$$

This rank-based procedure avoided direct averaging of log2 fold changes measured using different technologies. The resulting signatures, directions and gene orderings were fixed for discrete recovery, rank-weighted recovery, extreme-tail enrichment, pathway analyses and cross-omics comparisons.

**Direction-specific recovery and disease-strength-stratified analysis**

A predicted network effect was considered *disease-opposing* when its SRWR sign was negative for a molecule upregulated in STAD or positive for a molecule downregulated in STAD. For each molecular modality, disease-UP and disease-DOWN features were ranked separately by disease-effect magnitude. The predefined top fractions were $q \in \{5\%, 10\%, 20\%, 25\%, 50\%, 100\%\}$. For compound $c$, disease-UP suppression and disease-DOWN restoration were calculated as

$$\mathrm{REC}_{c,q}^{\mathrm{UP}} = \frac{1}{|D_{\mathrm{UP},q}|} \sum_{g \in D_{\mathrm{UP},q}} \mathbf{1}\left(Y_{cg} < 0\right),$$

$$\mathrm{REC}_{c,q}^{\mathrm{DOWN}} = \frac{1}{|D_{\mathrm{DOWN},q}|} \sum_{g \in D_{\mathrm{DOWN},q}} \mathbf{1}\left(Y_{cg} > 0\right),$$

where $Y_{cg}$ denotes the frozen signed SRWR influence score. An equally weighted macro-average of the two directions was defined for each compound:

$$\mathrm{REC}_{c,q}^{\mathrm{Combined}} = \frac{\mathrm{REC}_{c,q}^{\mathrm{UP}} + \mathrm{REC}_{c,q}^{\mathrm{DOWN}}}{2}.$$

The two directions were always retained separately because their behavior could differ, and the median of compound-level combined recovery values need not equal the average of the separately reported group medians. The top 10% and top 20% core disease-strength comparisons were displayed, and the complete set of six predefined fractions was retained for multiplicity-aware reporting.

As a continuous alternative to selecting a top fraction, rank-weighted recovery was calculated across all significant disease features. For a directional signature containing $N$ genes, a gene ranked $r$ received the normalized linear weight

$$w_r = \frac{N - r + 1}{N(N+1)/2}.$$

The corresponding direction-specific scores were

$$\mathrm{RWREC}_{c}^{\mathrm{UP}} = \sum_{g \in D_{\mathrm{UP}}} w_g \ \mathbf{1}\left(Y_{cg} < 0\right),$$

$$\mathrm{RWREC}_{c}^{\mathrm{DOWN}} = \sum_{g \in D_{\mathrm{DOWN}}} w_g \ \mathbf{1}\left(Y_{cg} > 0\right).$$

These statistics describe the orientation of inferred network influence with respect to the disease signature, not observed normalization of gene expression after treatment. Group-level differences were assessed using the fixed matched-control design described below.

**Extreme-tail enrichment of disease-opposing network influence**

To assess whether the strongest SRWR effects were concentrated on disease-associated molecular features, the entire 16,899-node network profile for each compound was sorted by signed influence score. The lowest and highest fractions of the profile defined negative and positive tails, respectively. Tail fractions of 2.5%, 5% and 10% were retained; 5% was the principal view, whereas the other fractions were sensitivity analyses. A negative-tail overlap with disease-UP features represented predicted suppression, and a positive-tail overlap with disease-DOWN features represented predicted restoration. These comparisons were performed separately for RNA, protein and the concordant core.

For a given tail and disease feature set, a 2-by-2 table enumerated the number $a$ of genes in both sets, $b$ in the tail only, $c$ in the disease set only, and $d$ in neither set. A 0.5 Haldane–Anscombe correction was applied to each cell before calculating the odds ratio:

$$\mathrm{OR} = \frac{(a + 0.5)(d + 0.5)}{(b + 0.5)(c + 0.5)}.$$

For the negative and positive SRWR tails, direction-specific log2 odds-ratio scores were defined as

$$\mathrm{M5}_{c}^{\mathrm{UP}}(q) = \log_2 \mathrm{OR}\left(T_{c}^{\mathrm{neg}}(q), D_{\mathrm{UP}}\right),$$

$$\mathrm{M5}_{c}^{\mathrm{DOWN}}(q) = \log_2 \mathrm{OR}\left(T_{c}^{\mathrm{pos}}(q), D_{\mathrm{DOWN}}\right).$$

The two directional log2 odds ratios were combined within each compound as

$$\mathrm{M5}_c^{\mathrm{Combined}}(q) = \frac{\mathrm{M5}_c^{\mathrm{UP}}(q) + \mathrm{M5}_c^{\mathrm{DOWN}}(q)}{2}.$$

Positive log2 odds ratios indicate absolute enrichment relative to the network background; a negative value indicates depletion even if it is more favorable than a matched-control value. Log2 scores were exponentiated back to the odds-ratio scale for presentation. The full tested grid and its multiple-testing results were retained, rather than choosing the best-performing signature or tail fraction after examination.

**Directional pathway reversal across curated gene-set collections**

Biological-program-level evaluation used the human MSigDB v2026.1 Hallmark, Reactome, KEGG MEDICUS and Gene Ontology Biological Process (GO-BP) collections. Each gene set was mapped to the common human gene identifiers and intersected with the significant disease signature and STAD network features. Eligibility was determined by the number of mapped disease-associated genes, not by prior enrichment significance: Hallmark required at least 10 disease genes, while Reactome, KEGG MEDICUS and GO-BP required 10–500. RNA, protein and concordant-core pathways were evaluated separately. The original catalogues and the final eligible-set memberships were preserved.

Within each omics-specific disease-UP and disease-DOWN group, genes were ranked by absolute disease effect and assigned linearly descending weights. The direction was encoded by the sign of the tumor-associated differential effect. For a gene $g$ at rank $r_g$ among $N$ genes in its directional set, the signed disease weight was

$$w_g^D = \mathrm{sign}(\Delta_g^D)\,\frac{N - r_g + 1}{N(N+1)/2}.$$

For each eligible pathway $P$, the weights of its disease-associated genes were renormalized to unit absolute mass:

$$w_{g,P}^D = \frac{w_g^D}{\sum_{h \in P \cap D} |w_h^D|}.$$

Separately, the full signed SRWR profile of each compound was transformed to within-compound network-wide rank scores $R_{cg}^{\mathrm{rank}}$ ranging approximately from −1 to +1, with the most negative and positive effects near the respective extremes. Directional pathway reversal was quantified as

$$\mathrm{PathRev}(c,P) = -\sum_{g \in P \cap D} w_{g,P}^D\, R_{cg}^{\mathrm{rank}}.$$

A positive score denotes an aggregate orientation opposite to the disease signature, whereas a negative score remains disease-concordant in absolute terms. In particular, a negative observed score that exceeds its matched-null counterpart indicates a relative shift *toward* reversal, not complete or absolute pathway reversal. Two compound-level summary metrics were calculated independently for each pathway catalogue and omics signature:

$$\mathrm{PathwayMedian}_c = \mathrm{median}_P[\mathrm{PathRev}(c,P)],$$

$$\mathrm{PathwayPosFrac}_c = \frac{\sum_P \mathbf{1}\,[\mathrm{PathRev}(c,P) > 0]}{N_{\text{eligible pathways}}}.$$

All four pathway catalogues and their aggregate outcomes, including nonsignificant results, were retained. Adjustments were calculated both within each six-test catalogue family and across the broader pathway-database comparisons. KEGG MEDICUS results were displayed as one view of the complete analysis, not as an independently validated replacement for the other catalogues.

**Cross-omics concordance of compound-associated molecular recovery**

To test whether individual compounds had relatively favorable disease-opposing scores at both the RNA and protein levels, each frozen RNA-specific and protein-specific recovery metric was rank-transformed independently over the same 1,682-compound evaluation universe. Average ranks were assigned to ties, and percentile scores were used instead of directly averaging quantities with potentially different empirical distributions. Denoting the percentile scores of compound $c$ by $p_c^R$ and $p_c^P$,

$$p_c^R = \text{PercentileRank}(S_c^{\text{RNA}}), \qquad p_c^P = \text{PercentileRank}(S_c^{\text{Protein}}).$$

Two complementary consensus operators were defined:

$$C_c^{\text{MIN}} = \min(p_c^R, p_c^P),$$
$$C_c^{\text{GM}} = \sqrt{p_c^R p_c^P}.$$

The minimum operator requires a compound to perform favorably in both layers, whereas the geometric mean permits limited compensation while rewarding joint performance. Both operators were applied separately to disease-UP and disease-DOWN rank-weighted recovery, with the corresponding combined 5% extreme-tail comparison retained as a complementary view. All six combinations were carried through the same matched-set testing framework. The disease-DOWN minimum-based rank-weighted consensus was retained as the principal subsequent cross-omics endpoint; the geometric mean was used as a sensitivity metric. Compound percentile ranks describe agreement of *predictive scores*, not experimentally observed RNA–protein concordance within the same treated sample.

**Evaluation against independent drug-perturbation proteomics**

Two drug-perturbation proteomics case studies were examined as orthogonal tests of predicted protein response direction. STO-609 was used as a pharmacological reference compound and was not treated as a natural product; tanshinone IIA was selected as a natural-product case with reported perturbation data in AGS gastric cancer cells. The two methods were evaluated on the same proteins in each case. The direct mechanism-of-action (MoA) baseline was derived from the unthresholded DTIAM activation and inhibition scores, whereas the network-based prediction used the frozen STAD tumor-specific SRWR profile. Thus, the direct baseline was *not* the thresholded signed DTI seed set used to initiate propagation. For each compound–protein pair, the direct directional score was

$$m_{cg} = A_{cg} - I_{cg},$$

and the propagated score was

$$Y_{cg} = R_{cg}^{\text{pos}} - R_{cg}^{\text{neg}}.$$

Experimental direction was determined from the measured drug-treated versus control protein abundance ratio:

$$d_g^{\text{exp}} = \text{sign}\left[\log_2\left(\frac{P_g^{\text{treated}}}{P_g^{\text{control}}}\right)\right].$$

After mapping protein identifiers to gene symbols, the directional agreement of either model was quantified on the common evaluable protein set $G$ as

$$\text{Agreement} = \frac{1}{|G|}\sum_{g \in G} \mathbf{1}\left[\text{sign}(\hat{d}_{cg}) = d_g^{\text{exp}}\right],$$

where $\hat{d}_{cg}$ is the direct MoA score or the SRWR score. For STO-609, the historical validation covered 7,111 common evaluable proteins and a 1,181-protein differentially abundant subset represented. The source study identifier, exact drug conditions and differential-protein selection rule for this subset require confirmation against the original validation script and were not inferred from the summary plots.

For tanshinone IIA, the published AGS perturbation study included iTRAQ proteomics deposited as ProteomeXchange PXD000998. Protein changes were assigned from the two treatment-to-control normalized reporter-ratio comparisons. A strict reference set required the two ratios to agree in direction, yielding 70 proteins, of which 68 could be matched to both prediction methods. On these 68 proteins, 28 directions agreed with the direct MoA baseline and 39 with SRWR:

$$\text{Agreement}_{\text{Direct}} = \frac{28}{68} = 41.18\%, \qquad \text{Agreement}_{\text{SRWR}} = \frac{39}{68} = 57.35\%.$$

The paired correctness table comprised 15 proteins predicted correctly by both methods, 13 by the direct baseline only, 24 by SRWR only and 16 by neither. An exact two-sided McNemar test evaluated the two discordant cells and yielded

$$P_{\text{McNemar,exact}} = 0.0989.$$

These analyses assess concordance with protein-abundance response directions, rather than direct binding, protein activity, mechanistic causality or therapeutic efficacy. Proteins measured in a single perturbation study were not treated as independent biological experiments for compound-level generalization. Inferences from the STO-609 case were likewise kept separate from conclusions concerning natural products.

**Extension from compounds to multicomponent herbal systems**

To investigate the extension of disease-specific influence modeling beyond individual compounds, standardized HRB identities were connected to their constituent CMP entities using the previously frozen herb–compound association atlas. Constituents for which valid STAD-specific network influence profiles were available were linked to the concordant RNA–protein disease signature. The analysis focused on the number of genes with a disease-opposing direction among the strongest 5% positive and negative network influence tails, and examined herb-level core recovery under different RNA and protein disease-effect thresholds.

This analysis was intended as an exploratory bridge between chemically resolved herb composition and disease-contextualized network influence. It should not be interpreted as evidence of herb efficacy, constituent bioavailability, concentration-dependent response or pharmacological synergy. The present source documentation establishes the source identities, eligible compound profiles and displayed recovery variables, but does not fully specify the implemented compound-to-herb aggregation operator, treatment of opposing compound signals, constituent weighting, missing-compound handling or exact herb-level recovery denominator. Because these choices can change the mathematical interpretation, no unverified aggregation equation was introduced here. The final reproducible version of this subsection must be completed directly from the frozen analysis code and its input/output audit; the displayed descriptive results should not be used to reverse-engineer an assumed averaging rule.

**Matched-set empirical inference and multiple-testing control**

The same frozen matched-set randomization design was used for the comparative molecular recovery, extreme-tail, pathway and cross-omics analyses. Let $L$ denote the 34 literature-supported compounds and $C_b$ the $b$-th matched control set, also containing 34 compounds. For a compound-level score $S_c$, the observed and null-set statistics were defined as

$$T_{\text{obs}} = \text{median}_{c \in L}(S_c),$$
$$T_b = \text{median}_{c \in C_b}(S_c), \qquad b = 1, \ldots, 10000.$$

For metrics in which larger values represented more favorable disease-opposing influence, the one-sided upper-tail empirical significance was

$$P_{\text{empirical}} = \frac{1 + \sum_{b=1}^{10000} \mathbf{1}\,(T_b \geq T_{\text{obs}})}{10001}.$$

The plus-one correction prevents reporting zero Monte Carlo $P$ values. All results retain the distinction between the observed literature-set median, the distribution of matched-set medians and the mean of the matched null distribution. Raw nominal $P$ values were supplemented by Benjamini–Hochberg correction within explicitly defined testing families. For the concordant-core discrete recovery analysis, the UP, DOWN and combined six-fraction families were assessed separately, with a further correction across the complete set of 18 comparisons. The six cross-omics consensus tests formed one family. Pathway aggregates were adjusted within each six-test pathway catalogue and, separately, across the broader pathway-database comparison family. Extreme-tail metrics were reported with their corresponding prespecified molecular-signature and tail-fraction families.

The STAD literature collection had been used during exploratory development and comparison of some evaluation metrics. Accordingly, matched-null results were treated as exploratory or method-development evidence, even where nominal or within-family adjusted significance thresholds were met. Shared compounds, shared disease signatures and the reuse of one matched-control assignment across several endpoints preclude treating these readouts as independent experimental validations. Negative tests and broader-family FDR results were retained. Confirmatory performance claims would require an independent, prospectively specified evaluation set or orthogonal experiments not used to choose the scoring rules.

**Author Contribution**

B.W., J.K., and B.Z. jointly conceived and designed the study, contributed to data collection, and critically reviewed and revised the manuscript. B.W. led the methodological development and drafted the original manuscript. All authors reviewed and approved the final version of the manuscript.

**Code availability**

The source code supporting this study has been deposited in a private GitHub repository and will be made publicly available upon completion of peer review. During the peer-review process, the code is available from the corresponding author upon reasonable request.

**Data availability**

The integrated herb–compound database developed in this study will be made publicly available upon completion of peer review. During the peer-review process, the database is available from the corresponding author upon reasonable request.

**Competing interests**

The authors declare no competing interests.